\documentclass[sigconf]{acmart}

\usepackage{balance}

\def\BibTeX{{\rm B\kern-.05em{\sc i\kern-.025em b}\kern-.08emT\kern-.1667em\lower.7ex\hbox{E}\kern-.125emX}}
    
\setcopyright{acmcopyright}

\copyrightyear{2019}
\acmYear{2019}
\acmConference[MM '19]{Proceedings of the 27th ACM International Conference on
Multimedia}{October 21--25, 2019}{Nice, France}
\acmBooktitle{Proceedings of the 27th ACM International Conference on Multimedia (MM '19),
October 21--25, 2019, Nice, France}
\acmPrice{15.00}
\acmDOI{10.1145/3343031.3350920}
\acmISBN{978-1-4503-6889-6/19/10}

\begin{document}
\fancyhead{}

%
\title{User-Aware Folk Popularity Rank: User-Popularity-Based \\ Tag Recommendation That Can Enhance Social Popularity}
\renewcommand{\shorttitle}{User-Aware Folk Popularity Rank}

%
\author{Xueting Wang}
\email{xt\_wang@hal.t.u-tokyo.ac.jp}
\affiliation{%
  \institution{The University of Tokyo}
  \streetaddress{7-3-1 Hongo, Bunkyo-ku}
  \city{Tokyo}
  \country{Japan}
}

\author{Yiwei Zhang}
\email{zhangyiwei@hal.t.u-tokyo.ac.jp}
\affiliation{%
  \institution{The University of Tokyo}
  \streetaddress{7-3-1 Hongo, Bunkyo-ku}
  \city{Tokyo}
  \country{Japan}
}

\author{Toshihiko Yamasaki}
\email{yamasaki@hal.t.u-tokyo.ac.jp}
\affiliation{%
  \institution{The University of Tokyo}
  \streetaddress{7-3-1 Hongo, Bunkyo-ku}
  \city{Tokyo}
  \country{Japan}
}

%

%
\begin{abstract}
In this paper we propose a method that can enhance the social popularity of a post (i.e., the number of views or likes) by recommending appropriate hash tags considering both content popularity and user popularity. A previous approach called FolkPopularityRank (FP-Rank) considered only the relationship among images, tags, and their popularity. However, the popularity of an image/video is strongly affected by who uploaded it. Therefore, we develop an algorithm that can incorporate user popularity and users' tag usage tendency into the FP-Rank algorithm. The experimental results using 60,000 training images with their accompanying tags and 1,000 test data, which were actually uploaded to a real social network service (SNS), show that, in ten days, our proposed algorithm can achieve 1.2 times more views than the FP-Rank algorithm. This technology would be critical to individual users and companies/brands who want to promote themselves in SNSs.

\end{abstract}

%
%


%

\maketitle

\section{Introduction}
Online sharing services such as Flickr, Instagram, and Facebook are becoming popular or even necessary for many people to share their daily generated contents, such as images and videos.
In these services, the number of views, comments, and favorites received after uploading indicate the popularity of the content.
In the following discussion, these measures are referred to as ``social popularity'' or ``social popularity scores.''
There are many studies on predicting social popularity scores of posted contents~\cite{huang2018random,meghawat2018multimodal,zhang2018user}. 
However, the prediction performance is still limited and there are seldom researches working on how to enhance social popularity.
Thus, it is still an important and challenging issue for both individuals and corporates who wish to enhance their social popularity scores as much as possible, as these scores reflect how much attention is paid to the content.

\begin{figure}
  \includegraphics[width=\linewidth]{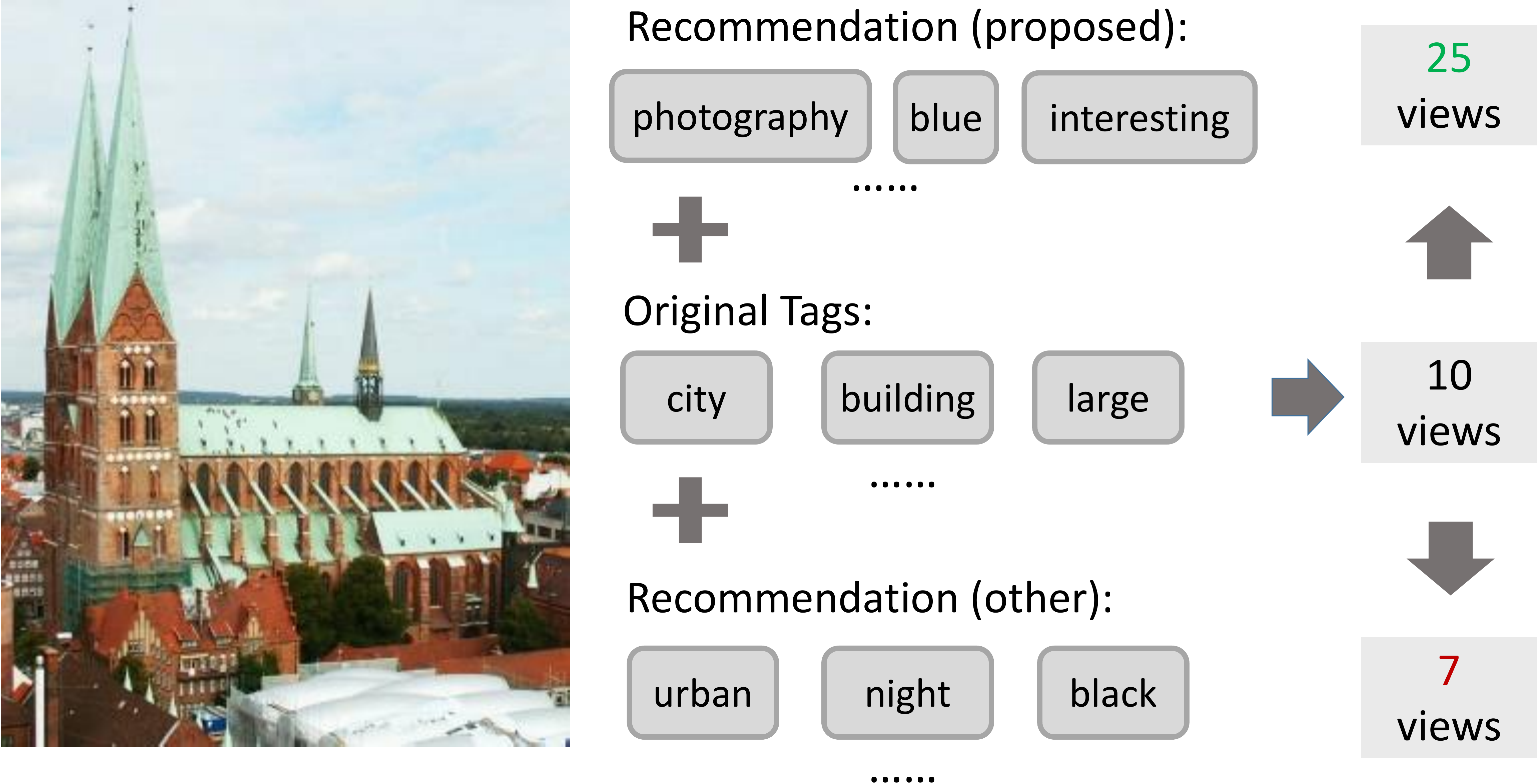}
  \caption{Example of social popularity change of different tag recommendation methods after uploading to the SNS in four days.}
  \label{fig:impression}
\end{figure}

We aim at enhancing social popularity of posted content using text tags attached to it.
Although the quality of the posted content is considered as an important factor, it has already been shown that the characteristics of the image or video are not effective in predicting the degree of social popularity~\cite{Hu2016,gelli2015image,yamaguchi2014chic}.
By contrast, in social networks, text tags attached to content play a very important role because users often search for content using text search engines.
Figure~\ref{fig:impression} shows an example of social popularity change in our uploading experiment (see Section~\ref{sec:experiment}), in which different tags with different recommendation methods result in different popularity for the same image in four days after uploading to SNS.
Therefore, it can be expected that social popularity enhancement can be achieved by ranking and recommending text tags based on their ability to affect social popularity.

Traditional tag ranking and recommendation systems in social media are often designed to recommend semantically relevant or collaborative-filtering-based tags~\cite{10.1007/978-3-319-57454-7_15,Yamasaki14,Lops13}.
However, we focus on methods of extracting tags that have influence on social popularity.
A tag recommendation algorithm for social popularity enhancement, called FolkPopularityRank (FP-Rank)~\cite{yamasaki2017folkpopularityrank}, was the first presented in a previous research.
The FP-Rank is inspired by the PageRank~\cite{ilprints422}, a well-known web page ranking algorithm, and the FolkRank~\cite{hotho2006folkrank}, an expanded version of PageRank considering the expanded network of users, contents, and tags. 
The FP-Rank can recommend tags to enhance social popularity based on the social popularity of the posted content and the co-occurrence among tags.
The popularity of a user (e.g., the sum of views or the number of followers), and the relation between users and tags is ignored.
However, in social networks, the social popularity of a certain content could be highly influenced by who upload it, especially when the user has high popularity. 

In this study, we propose a User-aware Folk Popularity Rank (UFP-Rank) algorithm, inspired by the FP-Rank, which ranks the tags of their social popularity influence by taking into account of not only the contents, but also the users.
Then, it combines tags by element-wise multiplication of two matrices to generate the appropriate adjacency matrix of tags weighted by the popularity of users, contents, and their co-occurrence with tags for increasing social popularity.
We use 60,000 training images with their accompanying tags and 1,000 test data, which were actually uploaded to a real social network service (SNS). 
The results showed that, in ten days, our proposed algorithm can achieve 1.2 times more views than the FP-Rank algorithm, and 2.8 times more views than the initial tags generated by an off-the-shelf computer vision API.
We can summarize our contributions as follows:
\begin{itemize}
\item Users' popularity information is verified to be effective for tag recommendation with popularity enhancement of posted contents via uploading experiments in real SNS.
\item Appropriate combination method of social popularity of users and contents with strong co-occurrence among tags is confirmed to be more effective for social popularity enhancement than existing methods, and achieved 2.8 times more views than the tags generated by an off-the-shelf computer vision API in SNS.
\item The proposed method performed effectively on recommending tags for content data in different domain with automatic generated tags other than user-annotated tags. In other words, the proposed method can be applied to tag recommendation with cold-start problems.
\end{itemize}

\section{Related Works}

\subsection{Graph-based Ranking}
\label{sec:graphrank}
We first introduce several algorithms which inspire our proposed method, the PageRank~\cite{ilprints422}, the FolkRank~\cite{hotho2006folkrank}, and the FP-Rank~\cite{yamasaki2017folkpopularityrank}. 

\subsubsection{PageRank}
The PageRank algorithm is an algorithm for ranking web pages developed by Google.
The basic concept is to rank web pages according to the link status between web pages.
Specifically, it consists of the following three hypotheses:
\begin {itemize}
    \item Web pages linked from many web pages are important.
    \item The more web pages a page links to, the less important each linked web page is.
    \item Web pages linked from important web pages are also important.
\end {itemize}
Thus, in the PageRank algorithm, web pages are regarded as a directed graph with pages as nodes and links as edges.
The edge weight is calculated by dividing the original web page's importance score by the number of links in the page.
To calculate this efficiently, the score of each node is calculated by repeating the equation~\eqref{Eq:pagerank} until convergence.
Then web pages can be ranked according to the obtained scores.
\begin{equation}
      \label{Eq:pagerank}
      \mathbf{r} = \alpha\mathbf{A}\mathbf{r}+(1-\alpha)\mathbf{p},
\end{equation}
where $\mathbf{r}$ is the importance score vector of each node, and $\mathbf{A}$ is the adjacency matrix of the web page graph model.
We use $\alpha$ as the damping factor to randomly consider the possibility of accessing other web pages.
$\mathbf{p}$ is a random surfer component.

\subsubsection{FolkRank}
FolkRank is an extended version of PageRank and is a ranking algorithm based on an undirected graph with links representing co-occurrence relationships among nodes of users, contents, and tags.
Here, the score of each node is calculated by changing the $ \mathbf{A} $ of tags into the adjacency matrix $ \mathbf{A}_{f} $ of users, contents, and tags.
As a result, content with important tags attached by important users is also treated as important.
This inference also applies to users and tags.
Furthermore, based on the existent tags, new tags are recommended by changing the preference vector $ \mathbf{p} $ as in the following Equation~\eqref{Eq:folkrank}.
\begin{equation}
\label{Eq:folkrank}
      \mathbf{w} = \mathbf{r}^{1} - \mathbf{r}^{0},
\end{equation}
where $ \mathbf{w} $ is a score vector of each tag.
For $ \mathbf{r}^{1} $, we set the existent tags to 1 and the others to 0 in the preference vector $ \mathbf{p} $.
$ \mathbf{r}^{0} $ gives equal weight to the preference vector $ \mathbf{p} $.
By setting the preference vector in this way, the tags co-occurring with the existent tags can be ranked high and extracted.

\subsubsection{FP-Rank}
FP-Rank is a tag recommendation algorithm, based on the concept of FolkRank for tag ranking and recommendation considering social popularity.
The adjacency matrix of tags not only reflects the co-occurrence among tags, but also adds the popularity of the contents as importance weights.
Consequently, tags attached to content with high social popularity are important, and the more tags are attached to a content, the less important each tag becomes.
To achieve this, $ \mathbf{A} $ is changed to the adjacency matrix of tags $ \mathbf{A}_{FP} $ weighted by the popularity of the posted contents.
The details and a comparison with the proposed method are described in Section~\ref{sec:proposed}.

\subsection{Tag Recommendation}\label{subsec:tag_recommendation}
Tag recommendation is a key technique for retrieval and navigation tasks in the web services, which usually depends on the information of both the target content information and the co-occurrence among users, tags, and contents.

Collaborative filtering (CF) is one of the most well-known recommendation techniques focused by researchers for a long time~\cite{resnick1994grouplens,konstan1997grouplens}.
Many CF-based methods are introduced in review papers~\cite{Su:2009:SCF:1592474.1722966,Lops13} and can be applied to tag recommendation using information such as tagging histories. Some researches made extension on collaborative filtering by matrix factorization~\cite{he2016fast} and neural-based approach~\cite{he2017neural}.

Tagcoor~\cite{Sigurbjornsson08} is a tag co-occurrence based method combining with ranking of user-based tag frequency. A tf-idf-like tag ranking algorithm was proposed using the tag frequency and weights learned from a regression model for each tag in~\cite{Yamasaki14}.

FolkRank~\cite{hotho2006folkrank} is another prominent technique that can be applied to tag ranking and recommendation as introduced in Section~\ref{sec:graphrank}.
Along with the basic FolkRank, numerous researches are conducted on improving FolkRank~\cite{Si09,Gemmell09,Zhang09,landia2012extending}.
Detailed information of the studies on collaborative tagging systems, also known as folksonomie can be found in a review paper~\cite{godoy2016folksonomy}.
Another survey paper compared the performance of different approaches of tagging systems focusing on adding semantics in folksonomies~\cite{Jabeen2016}.

Recent deep learning based techniques have also been applied for tag recommendation~\cite{10.1007/978-3-319-57454-7_15}.
A collaborative deep learning method jointly performs deep learning for the content information and CF for the ratings matrix~\cite{Wang:2015:CDL:2783258.2783273}.
Singhal et al. summarized recent recommendation systems using deep learning including collaborative systems, content-based systems and hybrid systems, where tag recommendation is one kind of application~\cite{10.5120/ijca2017916055}.

For most of these researches, tag recommendation and ranking are a kind of information retrieval task, which focus mainly on high co-occurrence or semantic precision.
In this study, in addition to these two points, we pay attention to the effect on social popularity of tags, which needs social popularity embedding mechanism.

\subsection{Social Popularity Prediction}\label{subsec:popularity_prediction}
Popularity prediction becomes a hot topic these years.
Popularity of online contents and user-generated-contents including news, products, and Youtube videos have been analyzed and predicted in~\cite{rathord2019comprehensive,rezaeenour2018developing,santosh2018product,yamaguchi2014chic,Nwana2013ALS}.
Along with the improvement of SNSs, social popularity prediction of the posted images~\cite{huang2018random,massip2018exploiting,lin2018layer,Wu:2016,Yamasaki14,Hu2016,McParlane:2014:NCH:2578726.2578776,Cappallo:2015:LFV:2671188.2749405,gelli2015image,vanZwol2010} and micro-videos~\cite{bielski2018pay,jing2018low,Chen:2016} are focused by both academics and industries.
Most of these popularity prediction are conducted by data-driven feature-based learning phases, and some researches embedded them with temporal models representing the variations on time information~\cite{santosh2018product,Wu:2016,DBLP:conf/icwsm/MishraRX18}.
Moreover, multi-modal features including textual and visual information, social connections have been investigated and combined with methods such as attention model to improve the predicting performance such as in~\cite{he2014predicting,meghawat2018multimodal,zhang2018user}.

However, \cite{Hu2016,gelli2015image,yamaguchi2014chic} showed that, in some services, the visual information have low predictive power compared to that of social cues. 
In addition, even deep learning based technique could achieve good performance, modeling visual features from billions of contents on SNSs suffers computational and cost problem in practice for most institutions and individuals.

For social connections, the dynamic nature of the connections themselves and cold start problem without connection histories of new users make this kind of feature difficult to be applied widely in practice.

Furthermore, context data such as tags attached with posted contents were most dominant in these works. 
In this study, instead of predicting the social popularity scores of existent contents on SNSs, we focus on how to enhance the level of popularity based on appropriate new tag recommendation.

\subsection{Tag Generation}\label{subsec:tag_generation}
There are many researchers working on automatic tag/text generation corresponding to visual contents such as images and videos.

Some researches have generated simple text descriptions for images using object-based features with traditional machine-learning methods~\cite{Yahong14,kulkarni2013babytalk,Chen:2016:SIP:2964284.2964306}.
Moreover, deep neural networks based approaches are investigated by researchers in many sub-fields such as image-text matching~\cite{10.1007/978-3-030-01225-0_13}, image captioning~\cite{gu2017stack}, and visual question answering~\cite{Anderson_2018_CVPR}.
A research in~\cite{chen2018deep} combined deep learning with traditional machine-learning by a new approach called ranking structural support vector machine with deep learning, which focused on the structural information among tags. 

For image tagging task, a work investigated image-word relevance relation in the word vector space and achieved zero-shot image annotation by approximating the principal direction from an image~\cite{zhang2016fast}. 
In~\cite{Wu2018TaggingLH}, they can annotate semantically relevant, distinct tags covering tags with diverse semantic levels of the image contents by sequential sampling from a determinantal point process model.

The generation results usually can semantically describe the objective facts in the contents, but lacking other related factors of tags, such as social effect on enhancing social popularity of the contents. 
 \begin{figure}[t]
        \centering
            \includegraphics[width=\linewidth]{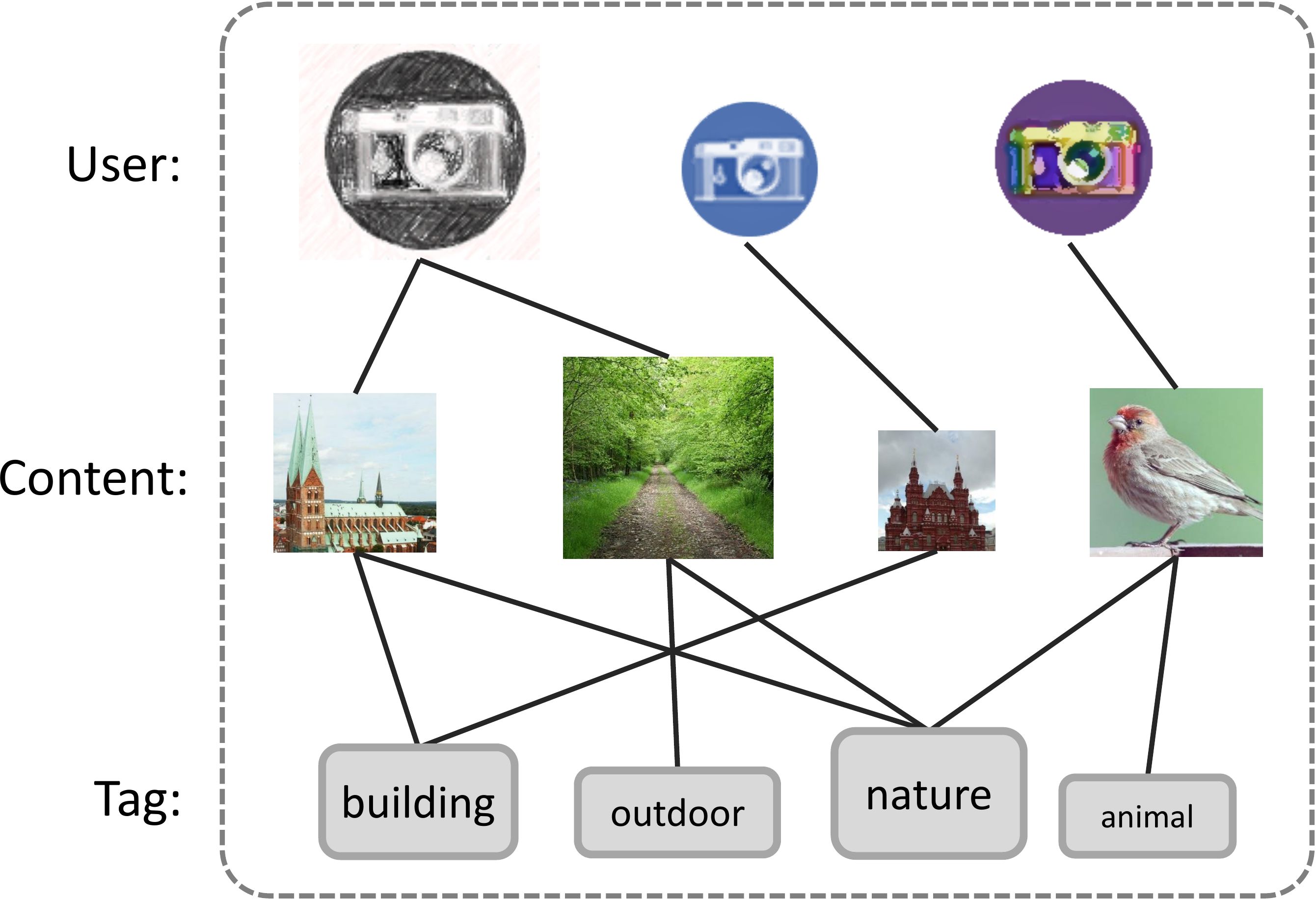}
          \caption{The concept of the tag recommendation using the relationship among users, contents, and tags for increasing social popularity.}
          \label{fig:concept}
    \end{figure}
     \begin{figure*}[t]
        \centering
            \includegraphics[width=\linewidth]{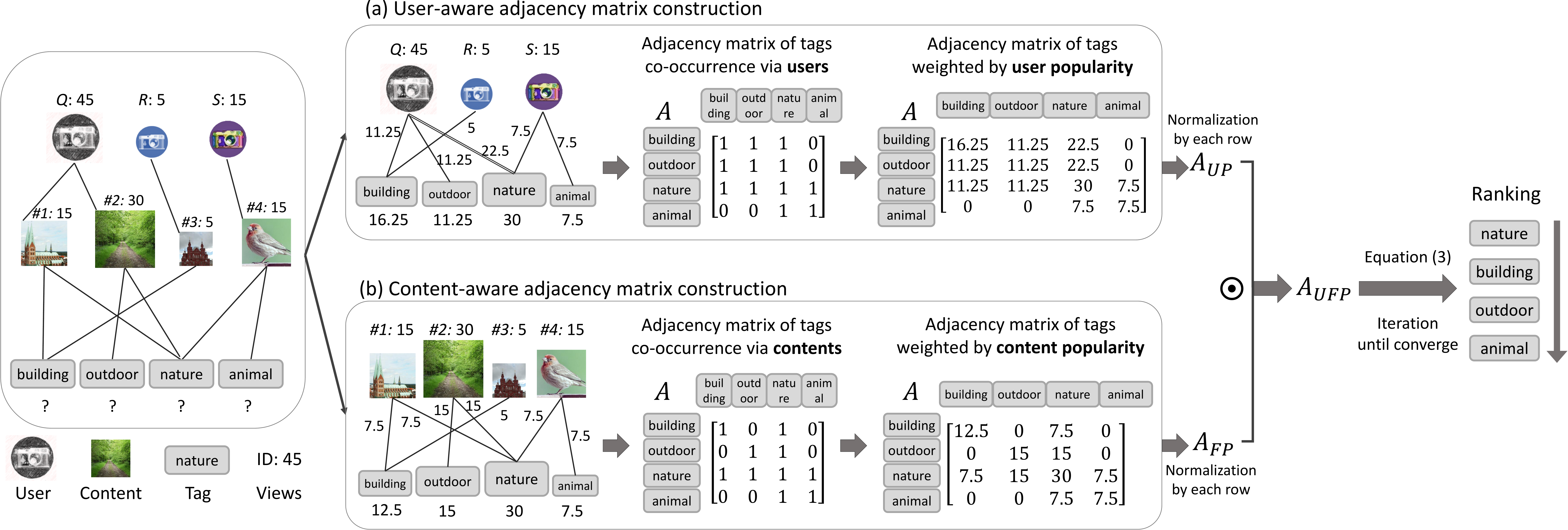}
          \caption{The outline of the proposed method: an example with visual interpretation of the adjacency matrix $ \mathbf {A}^{m}_{UFP} $ construction according to co-occurrence among tags and being weighted by popularity of users and contents for increasing social popularity.}
          \label{fig:outline}
    \end{figure*}
    
\subsection{User-aware Recommendation}\label{subsec:user_aware}
User context information including social relationship and online status/action/history, user attributes information including personality, emotion, and preference are applied in some researchers to improve the performance of recommendation tasks.

For tag recommendation, some researchers revised the user-based CF technique and proposed recommendation approaches fusing user-generated tags and social relations such as weighted friendship similarity in~\cite{ma2015social}.
Besides, there is a research using user tagging status to compute the tag probability distribution based on the statistical language model in order to recommend tags for users~\cite{yu2018tag}.

Moreover, researchers found that the experience quality of an application or a service is related to a user's personality~\cite{le2012qualinet}.
Some researches have embedded human attributes into model construction for experience assessment and video recommendation, such as personality and user preference~\cite{zhu2016qoe,Wang:2017:UGB:3132515.3132523}.

From these works, we can infer that user information is useful and critical in user-generated-content related tasks.
Thus, in this study, we would like to focus on social attributes of users and their effect on the social popularity.
Moreover, in these works, the user information usually used as features is necessary in both model training steps and test steps, which easily results in cold-start problems in practice.
While in our study, we only use user information in the training step to embed the trends of tag usage of users (especially users with high popularity), which can achieve effective tag recommendations to even new users without popularity information overcoming the cold-start problem. 

\section{User-Aware Folk Popularity Rank}

The proposed method performs tag recommendation considering the social popularity of not only posted contents but also the posting users in the source data.
As shown in Figure~\ref{fig:concept}, the proposed method can increase popularity by recommending tags according to the following concepts:
\begin{itemize}
\item Tags attached to contents with high social popularity are important tags.
\item The more tags attached to a content, the less important each tag is.
\item Tags that co-occur with important tags are also important.
\item Tags used by users with high social popularity are important.
\item The more tags used by a user, the less important each tag is, but tags used more frequently are more important.
\end{itemize}
The former three are based on the relationship among tags and posted contents, which are inspired from the concepts of PageRank (mentioned in Sec.2.1.1). 
As user popularity has not been investigated, we come up with the latter two concepts to include the relationship between tags and users. 

To achieve these, the proposed tag recommendation method consists of two steps: (1) a tag ranking step to calculate scores of tags by constructing a weighted adjacency matrix of tags from all posts in the source dataset; (2) a tag recommendation step to recommend new tags based on existent tags for posts in the target dataset.

    \subsection{Tag Ranking}
      \label{sec:proposed}
    \subsubsection{Tag Scoring}
    In this step, we calculate a score representing the ability on affecting social popularity of each tag. 
    Then the tags can be ranked based on the importance scores.
    The vector of scores of all tags $\mathbf{r}_{UFP}$ is calculated using weighted adjacency matrix of tags $\mathbf{A}_{UFP}$ by iterating from initial scores as follows:
      
      \begin{equation}
      \label{Eq:iterated}
      \mathbf{r}_{UFP} = \alpha\mathbf{A}_{UFP}\mathbf{r}_{UFP}+(1-\alpha)\mathbf{p},
      \end{equation}
    where $ \alpha $ is the damping factor set to 0.85 same with PageRank in this study, $\mathbf{p}$ is a preference vector (random surfer component) representing the importance of each tag.
    The weighted adjacency matrix of tags $\mathbf{A}_{UFP}$ is a square matrix with the size of $ T \times T $, $ T $ is the number of unique tags in the dataset.

    The initial vector of $\mathbf{r}_{UFP}$ can be set as equal values for all tags with the same sum with the preference vector. $\mathbf{r}_{UFP}$ can converge after approximately 50 iterations, and 10 iterations are sufficient for practical application as introduced in~\cite{ilprints422}.
    We confirmed this and set the converge standard as the same max iteration times or when the change is smaller than a threshold.
    
    \subsubsection{Adjacency Matrix Construction}
    The difference between the proposed method and FP-Rank is the design of the adjacency matrix of tags $\mathbf{A}_{UFP}$. 
    In FP-Rank~\cite{yamasaki2017folkpopularityrank}, the adjacency matrix of tags is calculated only considering the social popularity of posted contents represented as $\mathbf{A}_{FP}$.
    However, the popularity of a post is also affected by the user who upload it.
    Thus, we define an adjacency matrix of tags $\mathbf{A}_{UP}$ weighted by social popularity of users. 
    Here, $ \mathbf {A}_{FP} $ and $ \mathbf {A}_{UP} $ are the same size of $ T \times T $.
    In this study, we propose the following three kinds of matrix design using different features and combination methods to investigate their effectiveness on social popularity enhancement.
    \begin{itemize}
        \item \textbf{U-Rank}: $\mathbf{A}^{u}_{UFP}$ will be weighted only considering the association between user popularity and tags as shown in Equation~\eqref{Eq:user}.
        \item \textbf{UFP-plus-Rank}: element-wise addition of two matrices thus $\mathbf{A}^{p}_{UFP}$ will be weighted considering social popularity of both contents and users, while co-occurrence among tags exists when they are used by the same user as shown in Equation~\eqref{Eq:plus}.
        \item \textbf{UFP-product-Rank}: element-wise multiplication of two matrices thus $\textbf{A}^{m}_{UFP}$ will be weighted considering social popularity of both contents and users, while co-occurrence among tags only exists when they are attached to both the same user and the same posted content as shown in Equation~\eqref{Eq:product}.
    \end{itemize}
    \begin{figure}[t]
        \centering
            \includegraphics[width=\linewidth]{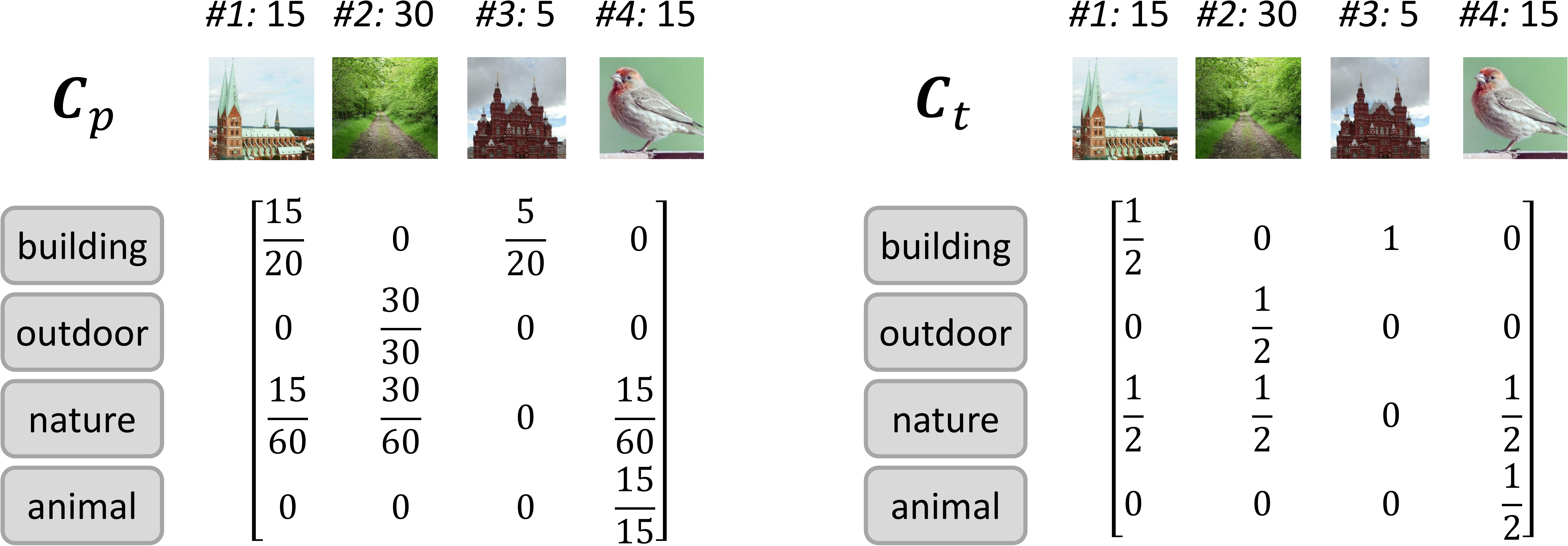}
          \caption{The sub matrices $\mathbf {C}_{p}$ and $\mathbf {C}_{t}$ for $ \mathbf{A}_{FP}$ calculation.}
          \label{fig:equation_fp}
    \end{figure}
    \begin{equation}
     \label{Eq:user}
        \mathbf{A}^{u}_{UFP} = \mathbf{A}_{UP},
    \end{equation}
    \begin{equation}
    \label{Eq:plus}
    \mathbf{A}^{p}_{UFP} = \mathbf{A}_{FP}+\mathbf{A}_{UP},
    \end{equation}
    \begin{equation}
    \label{Eq:product}
        \mathbf{A}^{m}_{UFP} = \mathbf{A}_{FP}\odot\mathbf{A}_{UP}.
    \end{equation}
    
    We show the outline of the proposed adjacency matrix construction with a visual interpretation first using UFP-product-Rank as a sample in Figure~\ref{fig:outline}.
    To easily explain the algorithm, we provide an example for a dataset including four images using four tags by three users, while user $Q$ uploaded two images \#1 and \#2, user $R$ and user $S$ uploaded image \#3 and \#4, respectively.
    Image \#1 is attached with tags \#building and \#nature, image \#2 attached with tags \#outdoor and \#nature, image \#3 with a tag \#building, and image \#4 with a tag \#animal.
    The social popularity scores (which can be applied by the numbers of views, comments, or favorites) of images \#1 to \#4 are 15, 30, 5, and 15, respectively.
    The social popularity scores of users (which can be calculated by the number of followers, the sum of views or favorites) are the sum of their posted images in this case: 45, 5, and 15.
    
    As shown in the figure, the $ \mathbf{A}_{UP} $ and $ \mathbf{A}_{FP} $ are calculated by the relation between users and tags, and relation between contents and tags, respectively.
    Here, we take $\mathbf{A}_{UP}$ construction (Figure~\ref{fig:outline}(a)) for an example. 
    First, the popularity scores are distributed from users to tags by considering the usage frequency of tags, such as for user $Q$, three tags have been used while \#nature is used twice. 
    Thus, tag \#nature is assigned $2/4$ of the score $45$ and each of the other two tags used by user $Q$ is assigned $1/4$.
    Secondly, in the adjacency matrix of tags $ \mathbf{A}$, the co-occurrence between two tags exits ($ > 0$) when they are used by the same user.
    The matrix can be weighted by the assigned popularity scores according to the co-occurrence among tags and users, such as co-occurrence exits between tags \#building and \#nature and the value of the two in the matrix can be weighted as $7.5$ because they are attached to the same content \#1.
    Consequently, we can get adjacency matrix of tags $\mathbf{A}_{UP}$ after normalization the weighted matrix by each row.
    Note that, the weights of popularity to \#building and \#outdoor in the matrices are different in $\mathbf{A}_{UP}$ and $\mathbf{A}_{FP}$, which could result in different ranking results when only use U-Rank or FP-Rank in this example.
    
    Then, we describe the concrete calculation of the adjacency matrix of tags.
    The matrices $ \mathbf{A}_{FP} $ and $ \mathbf{A}_{UP} $ can be calculated as a combination of two sub matrices, which representing the relation between tags and contents and the relation between tags and users as follows:
    \begin{equation}
      \label{Eq:fp}
      \mathbf{A}_{FP} = \mathbf{C}_{p} \times \mathbf{C}_{t}^{T},
    \end{equation}
    \begin{equation}
      \label{Eq:up}
      \mathbf{A}_{UP} = \mathbf{U}_{p} \times \mathbf{U}_{t}^{T}.
    \end{equation}
     
    First, we introduce the construction of $ \mathbf{A}_{FP}$ used in FP-Rank.
    $ \mathbf{C}_{p} $ and $ \mathbf{C}_{t} $ are matrices with the size of $ T \times I $, where $ I $ is the number of contents in the source dataset.
    The $ i $th row vector of $ \mathbf{C}_{p} $ is a set of social popularity scores of contents attached with tag $ i $, normalized by the sum of the scores in the row.
    The $ j $th column vector of $ \mathbf{C}_{t} $ is the usage flag for each tag, normalized by the total number of tags attached to $j$th content.
    In the example dataset in Figure~\ref{fig:outline}, 
    $\mathbf {C}_{p}$ and $\mathbf {C}_{t}$ can be calculated as shown in Figure~\ref{fig:equation_fp}.

    Consequently, the element $ a ^ {C}_{ij} $ of $ \mathbf{A}_{FP} $ is calculated as follows:
          \begin{equation}
          a^{C}_{ij} = \sum_{d\in \mathbf{D}}\frac{u^{C}(d)}{(\mbox{No. of tags attached to content $ d $})},
          \end{equation}
          
          \begin{align}
           u^{C}(d) = \frac{(\mbox{social popularity of content $d$})+k}{\sum{(\mbox{social popularity of the contents with tag $i$})}},
          \end{align}
    where $ d $ is the index of the content attached with tags $ i $ and $ j $ simultaneously, $\mathbf{D}$ is a set of content in the source data, $ u^{C}(d) $ is the weight of tags $ i $ and $ j $ calculated by the social popularity of content $ d $. $ k $ is a parameter to prevent $ u^{C} (d) $ from becoming $ 0 $.

\begin{figure}[t]
        \centering
            \includegraphics[width=\linewidth]{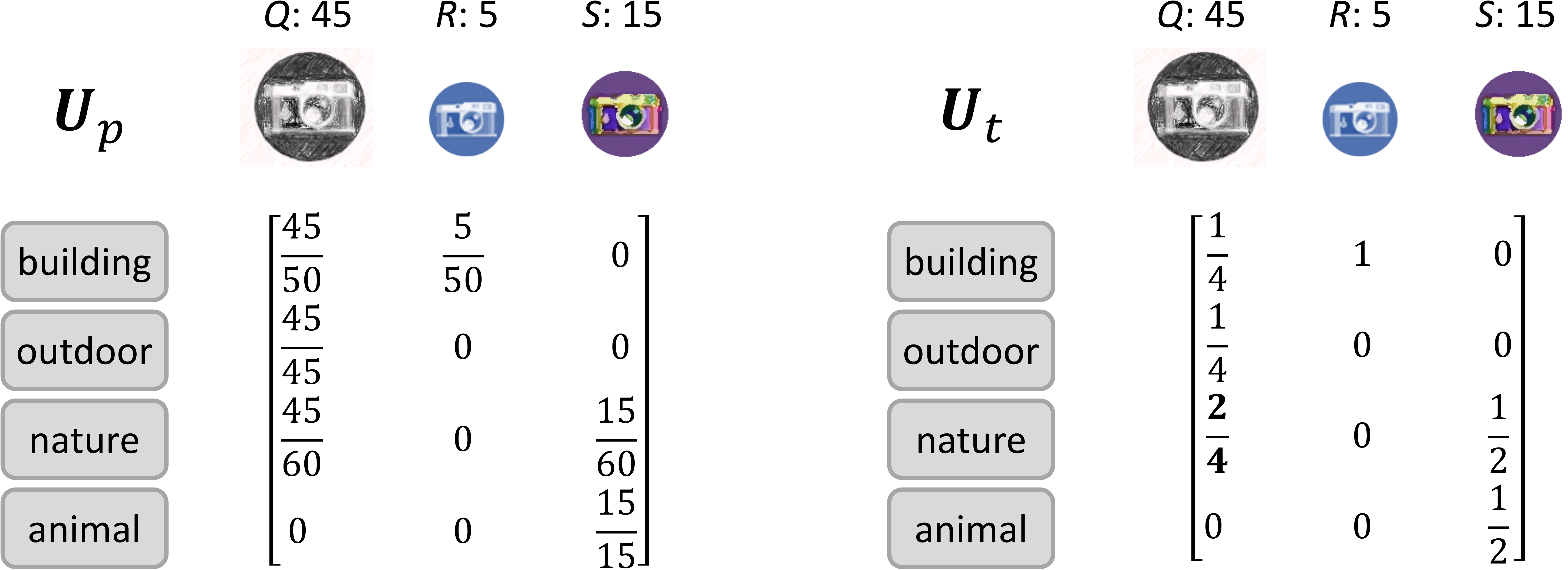}
          \caption{The sub matrices $\mathbf {U}_{p}$ and $\mathbf {U}_{t}$ for $ \mathbf{A}_{UP}$ calculation.}
          \label{fig:equation_up}
    \end{figure}

    Different to FP-Rank ($ \mathbf {A}_{FP} $), $ \mathbf {A}_{UP} $ is calculated using the users' social popularity and tag usage frequency according to Equation~\ref{Eq:up}.
    $ \mathbf {U}_{p} $ and $ \mathbf {U}_{t} $ are $ T \times N $ sized matrices, $ N $ is the number of users in the source dataset.
    The $ i $th row vector of $ \mathbf {U}_{p} $ is a set of users' social popularity scores using the tag $ i $, normalized by the sum of scores in the row.
    In this paper, the social popularity of a user is the total number of social popularity of his/her posted contents.
    Since the user can use a tag multiple times, the $j$th column vector of $\mathbf{U}_{t}$ is the frequency of each tag has been used normalized by the total usage frequency of tags used by the $j$th user. 
    Using the example dataset in Figure~\ref{fig:outline}, $\mathbf {U}_{p}$ and $\mathbf {U}_{t}$ be calculated as shown in Figure~\ref{fig:equation_up}.

    Consequently, the element $ a^{U}_{ij} $ of $ \mathbf{A}_{UP} $ is calculated as follows:
          \begin{equation}
          a^{U}_{ij} = \sum_{l\in \mathbf{L}}\frac{u^{U}(l)\times (\mbox{the usage frequency of tag $j$ by user $l$}) }{(\mbox{the sum of usage frequency of all tags by user $l$})},
          \end{equation}
          \begin{align}
          u^{U}(l) = \frac{(\mbox{social popularity of user $l$})+k}{\sum{(\mbox{social popularity of users using tag $i$})}},
          \end{align}
    where $ l $ is the index of the user simultaneously using the tags $ i $ and $ j $, $ \mathbf {L} $ is the set of users in the source dataset. $ u ^ {U} (l) $ is the weight of tag $ i $ and $ j $ calculated by the popularity of user $ l $. $ k $ is a parameter to prevent $ u^{U} (l) $ from becoming $ 0 $.

    \subsection{Tag Recommendation}
    Similar to the FolkRank, based on the tags already attached to the content, we can recommend new tags through the following equation:
    \begin{equation}
          \mathbf{w}_{UFP} = \mathbf{r}_{UFP}^{1} - \mathbf{r}_{UFP}^{0},
    \end{equation}
    where $ \mathbf {w}_{UFP} $ is a vector of the final ranking score of all tags considering the co-occurrence with existent tags.
    The difference between $ \mathbf {r}_{UFP} ^ {1} $ and $ \mathbf {r}_{UFP} ^ {0} $ is the setting of the preference vector $ \mathbf {p} $ in Equation~\eqref{Eq:iterated} when calculating $ \mathbf {r}_{UFP}$ for the target dataset using the constructed matrix $\mathbf {A}_{UFP}$. 
    For example, when we generate $ \mathbf {r}_{UFP} ^ {1} $, the tags already attached to the post are weighted as 1, and the others have a weight of 0 in the preference vector $ \mathbf {p} $.
    For $ \mathbf {r} _ {UFP} ^ {0} $, we give equal weights to all tags in the preference vector and the sum of weights is the same as in the preference vector of $ \mathbf {r}_{UFP} ^ {1} $.
    By setting the preference vector in this way, tags that co-occur with tags already attached can be extracted.
    
    Both $ \mathbf {r}_{UFP} ^ {1} $ and $ \mathbf {r}_{UFP} ^ {0} $ are iterated until convergence.
    The resulting scores of $ \mathbf {w}_{UFP} $ reflect the co-occurrence with the tags already attached, and their influence on the social popularity.
    Tags are ranked according to these scores and the top tags are recommended as new tags.

\section{Experiments}
    \label{sec:experiment}
    \subsection{Dataset}
        In this study, the number of views is used as the measure of social popularity.
        For the source dataset to train the adjacency matrix of tags, we randomly select 60,000 images (uploaded by 6462 users) with over 20 tags and over 5000 views from Flickr's public data set YFCC100M~\cite{thomee2016yfcc100m}.
        Consequently, there are over 254,000 unique tags used in popular posts on the SNS included in our dataset, which is a broad resource for constructing a generalized matrix of tags.
        More details can be found in Table~\ref{table:dataset}.
        
        For testing, in the target dataset, contents with annotated initial tags are needed for new tags recommendation.
        However, some users prefer not or unable to annotate appropriate tags by themselves before automatic recommendation. 
        Thus, regarding the cold-start problem in practical recommendation, we created a target dataset including 1000 images randomly selected from Wikimedia Commons\footnote{https://commons.wikimedia.org/wiki/Main\_Page} for testing.
        And then corresponding initial tags were generated according to the image contents automatically by a computer vision API provided by the Microsoft Cognitive Services\footnote{https://azure.microsoft.com/en-us/services/cognitive-services/computer-vision/} (MCS). 
        The detail of the dataset can be found here~\footnote{https://github.com/xueting-wang/UFP-Rank}.  
        
        \begin{table}[t]
            \begin{center}
            \caption{Overview of the source and target datasets.} 
            \label{table:dataset}
            \begin{tabular}{|c|c|c|}
            \hline 
            Data Set &  Source & Target\\ \hline \hline
            Total Number of Images & 60,000 & 1000 \\ \hline
            Average Number of Views of an Image&  13,139.5 & - \\ \hline
            Average Number of Tags of an Image & 37.1 & 23.1 \\ \hline
            Total Number of Users & 6462 & - \\ \hline
            Average Number of Images of a User &  9.3 & - \\ \hline
            Average Number of Views of a User & 122001 & - \\ \hline
            \end{tabular}
            \end{center}
        \end{table}
        
    \begin{figure*}[t]
        \centering
            \includegraphics[width=\linewidth]{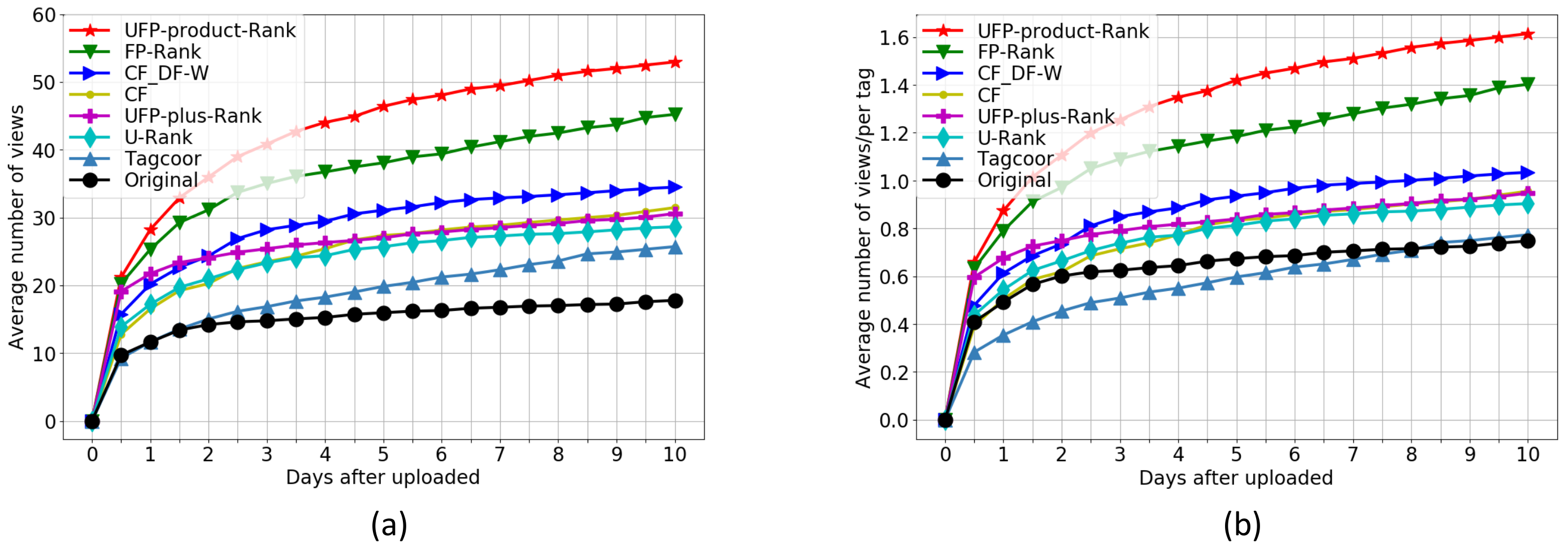}
          \caption{(a) Popularity change on the average number of views of each image with different methods lasting for ten days after uploading; (b) Popularity change on the average number of views of each image and each tag with different methods.}
          \label{fig:evaluation_views}
        \end{figure*}    
        
    \subsection{Comparative Methods}
        In the experiment, we recommended the top 10 tags, different from the initial ones, according to the ranking results of different methods.
        
        First, we evaluate the recommendation performance of the proposed methods: 
        (1) U-Rank ($ \mathbf{A}^{u}_{UFP}$), considering only the relation between users and tags; 
        (2) UFP-plus-Rank ($\mathbf{A}^{p}_{UFP}$) and
        (3) UFP-product-Rank ($\mathbf {A}^{m}_{UFP}$), considering the association among tags, contents, and users.
        
        Then, we compare the proposed methods with five other recommendation methods: 
        (4) Original (MCS), the tags generated by an off-the-shelf computer vision API without recommendation; 
        (5) Tagcoor~\cite{Sigurbjornsson08};
        (6) Collaborative Filtering (CF)~\cite{Su:2009:SCF:1592474.1722966,Lops13};
        (7) CF with DF-W (CF\_DF-W)~\cite{Yamasaki14},
        and 
        (8) FP-Rank ($ \mathbf {A}_{FP}$)~\cite{yamasaki2017folkpopularityrank}, considering only the relation between contents and tags.
        All the algorithms were implemented by ourselves.
        
        Because FP-Rank has been introduced in detail in Section~\ref{sec:proposed}, we briefly describe the last three comparative methods.
        Tagcoor makes recommendations based on tag co-occurrences and is defined as follows:
        \begin{equation}
        P(t_j \mid t_i) = \frac{\mid t_i \cap t_j \mid}{\mid t_i \mid} 
        \end{equation}
        Tag aggregation and promotion strategies are then used to produce the final list of recommended tags.
        The idea of CF for tag recommendation is to suggest new tags based on annotations of similar images by a collaborative filtering phase to generate candidate rand ranking phase to rank them.
        Given an image with original tags, a feature vector is represented by the set of tags, which is defined as follows:
        \begin{equation}
        FV_{i}^{tag} = \{f_{i1}, f_{i2}, \cdots, f_{ij}, \cdots,f_{iT} \},
        \end{equation}
        where $FV_{i}^{tag}$ is the feature vector for the $i$th image tag sets, $f_{ij}$ indicates whether the $i$th image has the $j$th tag. 
        These vectors can be viewed as approximate representations of the image content and user preference in the corresponding domain. An ordered list of candidate tags is derived based on the similarity between tag sets, and the similarity is measured by computing the cosine of the angle formed by the two feature vectors. 
        Then, the candidate tags are ranked and the top $n$ tags in the ranking list will be recommended to the users. 
        The basic CF method ranks the candidate tags by their frequency to ultimately produce the ranked list of recommended tags. 
        For the CF\_DF-W (document frequency-weights from regression) method, a linear SVR model is trained by using the feature vector and the social popularity scores as target value to obtain the weight vector for ranking candidate tags.

    \subsection{Evaluation by Uploading to Flickr}
        
        To acquire a relevant evaluation of the effect of increasing social popularity, we uploaded the recommendation results on Flickr and investigated the changes in social popularity (the number of views).
        We uploaded the results eight times for all the proposed and comparative methods one by one to avoid multiple identical images with tags recommended by different methods to be seen at the same.  
        We created a new Flickr account for test each time to avoid history effects. 
        Note that, the user-aware recommendation model was constructed based on users' information in a relatively large training dataset, thus in the test we can recommend effective tags even for a new user account.
        Each time, we uploaded testing images with automatically generated initial tags and 10 recommended tags. 
        Then, we checked the number of views twice a day (every 12 hours, which did not affect the number of views), and deleted all the files and accounts after ten days. 
        Therefore, the evaluation experiments of different methods are ensured to be done independently. 
        Moreover, we conducted each uploading almost in the same time of a day and avoided special days such as the new year to reduce the influence of periods.
        The whole uploading experiment lasted from Sep. 2018 to Feb. 2019.
              
        \begin{figure}[t]
         \centering
         \includegraphics[width=\linewidth]{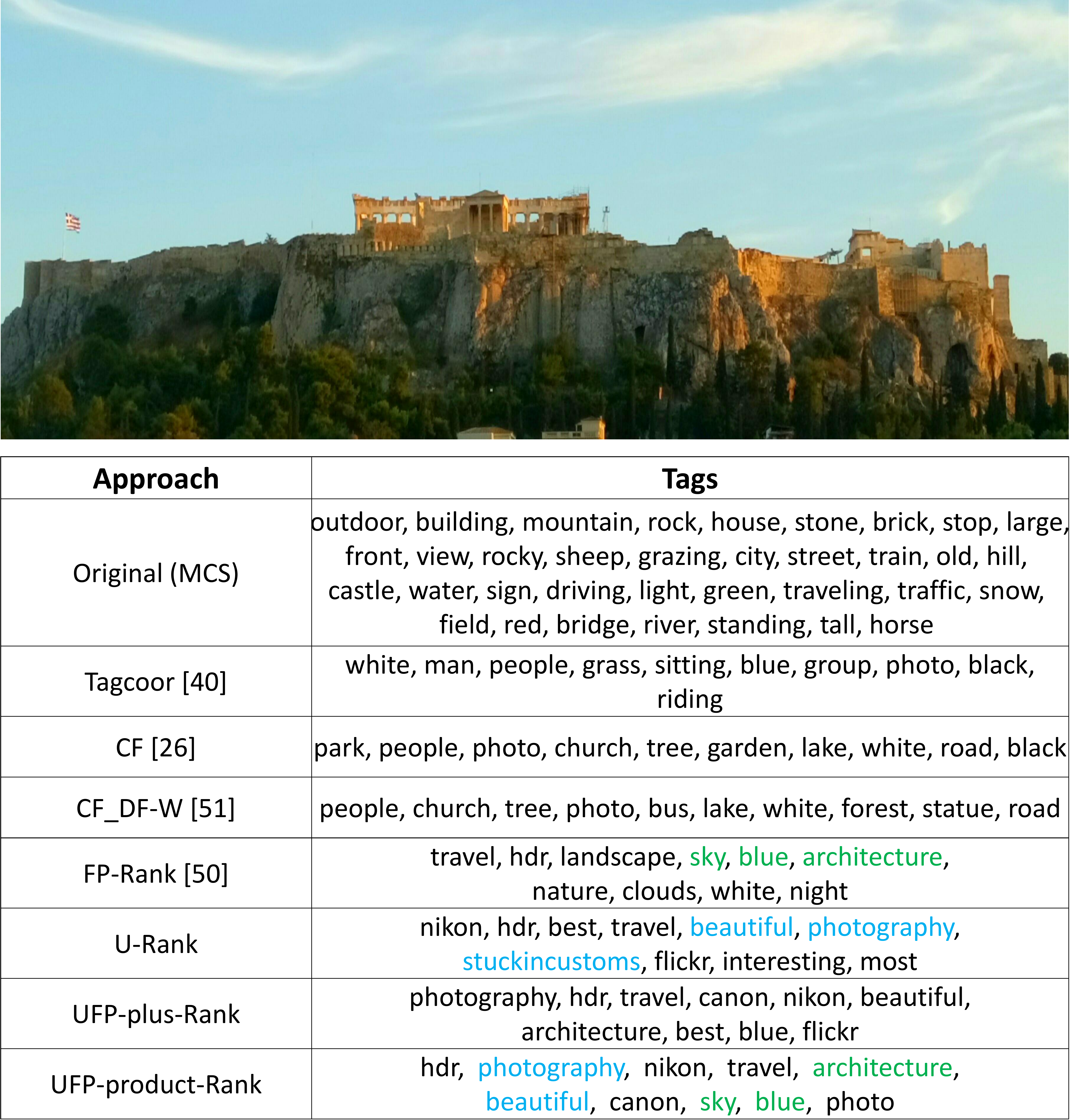}
        	\caption{Examples of the recommendation result.}
        	\label{fig:result_examples}
         \end{figure}
    
\section{Results and Analysis}
    First, we quantitatively discuss the effect of increasing the popularity of recommendation tags based on different methods by the uploading experiments.
    The Figure~\ref{fig:evaluation_views}(a) shows the average number of views of each image with each method for the test data.
    We can see that the UFP-product-Rank achieved the highest number of views when uploaded after ten days. 
    The number of views of UFP-product-Rank is almost 2.8 times larger than that of the initial tags generated by a computer vision API provided by the Microsoft Cognitive Services (Original) and 1.2 times larger than (significantly higher by paired T-test, $p < 0.01$) that of the results of FP-Rank, which was the highest comparative method.
    In addition, all the user-aware proposed methods (UFP-product-Rank, UFP-plus-Rank, U-Rank) can improve social popularity from just using initial tags.
    These results verified the effectiveness of using users' social popularity for popularity enhancement in SNSs.
    
    By contrast, the recommendation only using users' social popularity (U-Rank) and weak co-occurrence among tags as long as they are used by the same user (UFP-plus-Rank), improved less than the content-aware FP-Rank and co-occurrence based CF\_DF-W.
    It can be considered that the co-occurrence among tags in the U-Rank and UFP-plus-Rank is weakened from ``attached to the same image'' to ``can be attached to different images as long as posted by the same user.''
    Thus, some tags with less positive or even negative effects on popularity may be involved comparing to the FP-rank and CF\_DF-W.
    Consequently, it can be inferred that the appropriate combination of popularity of users and contents with strong co-occurrence among tags is important for social popularity enhancement.
    
    To avoid the influence of the number of tags (10 more than the number of initial tags for each recommendation method), we divided the average number of views of each image by the number of tags with each method.
    The result is shown in Figure~\ref{fig:evaluation_views}(b).
    We can find that the UFP-product-Rank still achieved the highest number of views over the other tag recommendation methods.
    However, for the method Tagcoor, the number of views has not changed a lot after ten days uploading and even lower than the original ones within 7 days after uploading.
    We can infer that Tagcoor might recommend some negative tags on the influence of social popularity, and it also validate that the number of tags is not the critical factor for social popularity enhancing.
    
    Then, we analyze the tag recommendations of different methods in detail.
    Figure~\ref{fig:result_examples} shows the examples of posted images with top 10 recommended tags with each method.
    Comparing the recommendation results in Figure~\ref{fig:result_examples}, 
    it can be seen that the U-Rank and UFP-Rank recommended more adjective tags (e.g., beautiful) and more tags expressing impression (e.g., photography) or association (e.g., Stuck In Customs, a travel photography blog) than the FP-Rank. 
    By contrast, the FP-Rank recommended more tags representing contents and objects (e.g., sky, blue), and better maintained the co-occurrence among tags.
    The CF-based methods have the similar trends with FP-Rank. 
    Consequently, users' social popularity is useful and meaningful for tag recommendation on social popularity enhancement in SNS.
    The appropriateness of the recommended tags by FP-Rank is already shown as good as human generated tags in~\cite{yamasaki2017folkpopularityrank}, and ours is an extension of the FP-Rank.
    As our proposed approach is based on not only the popularity but also the co-occurrence relation among tags, the recommended tags can increase popularity while keeping appropriateness as they tend to be related to the posted contents as the initial tags. 
    
\section{Conclusion}
    In this study, we proposed an algorithm that ranks and recommends tags according to their influence on social popularity (such as the number of views) in SNSs.
    The proposed method can increase the social popularity by considering the user's social popularity and the popularity of the posted content, along with the co-occurrence relation among tags.
    Especially, the element-wise multiplication of two matrices of user popularity and content popularity is the most effective way of generating the adjacency matrix of tags for recommendation.
    In addition, the proposed method can be applied to not only tag ranking and recommendation, but also user or content ranking and recommendation, similar to the FolkRank.
    This technology could benefit both individual users and companies/brands who want to promote themselves in SNSs.
    
    In the future, we plan to conduct experiments on more dataset and other SNSs to investigate the performance of the proposed method.
    We will also further consider how to support the creation or design of posting content to achieve more semantically relevant with higher degree of social popularity.

\begin{acks}
This is a joint project between CyberBuzz, Inc. and Yamasaki lab of The University of Tokyo. 
This work was partially financially supported by the Grants-in-Aid for Scientific Research Numbers JP19K20289, JP19J22939 and JP18H03339 from JSPS, and by JST-CREST (No. JPMJCR1686).
\end{acks}

%
\bibliographystyle{ACM-Reference-Format}
\balance
\bibliography{mm19.bib}


\begin{thebibliography}{56}


\ifx \showCODEN    \undefined \def \showCODEN     #1{\unskip}     \fi
\ifx \showDOI      \undefined \def \showDOI       #1{#1}\fi
\ifx \showISBNx    \undefined \def \showISBNx     #1{\unskip}     \fi
\ifx \showISBNxiii \undefined \def \showISBNxiii  #1{\unskip}     \fi
\ifx \showISSN     \undefined \def \showISSN      #1{\unskip}     \fi
\ifx \showLCCN     \undefined \def \showLCCN      #1{\unskip}     \fi
\ifx \shownote     \undefined \def \shownote      #1{#1}          \fi
\ifx \showarticletitle \undefined \def \showarticletitle #1{#1}   \fi
\ifx \showURL      \undefined \def \showURL       {\relax}        \fi
\providecommand\bibfield[2]{#2}
\providecommand\bibinfo[2]{#2}
\providecommand\natexlab[1]{#1}
\providecommand\showeprint[2][]{arXiv:#2}

\bibitem[\protect\citeauthoryear{Anderson, He, Buehler, Teney, Johnson, Gould,
  and Zhang}{Anderson et~al\mbox{.}}{2018}]%
        {Anderson_2018_CVPR}
\bibfield{author}{\bibinfo{person}{Peter Anderson}, \bibinfo{person}{Xiaodong
  He}, \bibinfo{person}{Chris Buehler}, \bibinfo{person}{Damien Teney},
  \bibinfo{person}{Mark Johnson}, \bibinfo{person}{Stephen Gould}, {and}
  \bibinfo{person}{Lei Zhang}.} \bibinfo{year}{2018}\natexlab{}.
\newblock \showarticletitle{Bottom-Up and Top-Down Attention for Image
  Captioning and Visual Question Answering}. In
  \bibinfo{booktitle}{\emph{Proceedings of the IEEE Conference on Computer
  Vision and Pattern Recognition (CVPR)}}. \bibinfo{pages}{6077--6086}.
\newblock


\bibitem[\protect\citeauthoryear{Bielski and Trzcinski}{Bielski and
  Trzcinski}{2018}]%
        {bielski2018pay}
\bibfield{author}{\bibinfo{person}{Adam Bielski} {and} \bibinfo{person}{Tomasz
  Trzcinski}.} \bibinfo{year}{2018}\natexlab{}.
\newblock \showarticletitle{Pay Attention to Virality: understanding popularity
  of social media videos with the attention mechanism}. In
  \bibinfo{booktitle}{\emph{Proceedings of the IEEE Conference on Computer
  Vision and Pattern Recognition Workshops (CVPRW)}}.
  \bibinfo{pages}{2335--2337}.
\newblock


\bibitem[\protect\citeauthoryear{Cappallo, Mensink, and Snoek}{Cappallo
  et~al\mbox{.}}{2015}]%
        {Cappallo:2015:LFV:2671188.2749405}
\bibfield{author}{\bibinfo{person}{Spencer Cappallo}, \bibinfo{person}{Thomas
  Mensink}, {and} \bibinfo{person}{Cees~G.M. Snoek}.}
  \bibinfo{year}{2015}\natexlab{}.
\newblock \showarticletitle{Latent factors of visual popularity prediction}. In
  \bibinfo{booktitle}{\emph{Proceedings of the ACM International Conference on
  Multimedia Retrieval (ICMR)}}. \bibinfo{pages}{195--202}.
\newblock


\bibitem[\protect\citeauthoryear{Chen, Xu, and Yang}{Chen
  et~al\mbox{.}}{2018}]%
        {chen2018deep}
\bibfield{author}{\bibinfo{person}{Gang Chen}, \bibinfo{person}{Ran Xu}, {and}
  \bibinfo{person}{Zhi Yang}.} \bibinfo{year}{2018}\natexlab{}.
\newblock \showarticletitle{Deep ranking structural support vector machine for
  image tagging}.
\newblock \bibinfo{journal}{\emph{Pattern Recognition Letters}}
  \bibinfo{volume}{105} (\bibinfo{year}{2018}), \bibinfo{pages}{30--38}.
\newblock


\bibitem[\protect\citeauthoryear{Chen, Jin, and Xiong}{Chen
  et~al\mbox{.}}{2016a}]%
        {Chen:2016:SIP:2964284.2964306}
\bibfield{author}{\bibinfo{person}{Jia Chen}, \bibinfo{person}{Qin Jin}, {and}
  \bibinfo{person}{Yifan Xiong}.} \bibinfo{year}{2016}\natexlab{a}.
\newblock \showarticletitle{Semantic Image Profiling for Historic Events:
  Linking Images to Phrases}. In \bibinfo{booktitle}{\emph{Proceedings of the
  24th ACM International Conference on Multimedia (ACMMM)}}.
  \bibinfo{pages}{1028--1037}.
\newblock


\bibitem[\protect\citeauthoryear{Chen, Song, Nie, Wang, Zhang, and Chua}{Chen
  et~al\mbox{.}}{2016b}]%
        {Chen:2016}
\bibfield{author}{\bibinfo{person}{Jingyuan Chen}, \bibinfo{person}{Xuemeng
  Song}, \bibinfo{person}{Liqiang Nie}, \bibinfo{person}{Xiang Wang},
  \bibinfo{person}{Hanwang Zhang}, {and} \bibinfo{person}{Tat-Seng Chua}.}
  \bibinfo{year}{2016}\natexlab{b}.
\newblock \showarticletitle{Micro Tells Macro: predicting the popularity of
  micro-Videos via a transductive model}. In
  \bibinfo{booktitle}{\emph{Proceedings of the ACM International Conference on
  Multimedia (ACMMM)}}. \bibinfo{pages}{898--907}.
\newblock


\bibitem[\protect\citeauthoryear{Gelli, Uricchio, Bertini, Del~Bimbo, and
  Chang}{Gelli et~al\mbox{.}}{2015}]%
        {gelli2015image}
\bibfield{author}{\bibinfo{person}{Francesco Gelli}, \bibinfo{person}{Tiberio
  Uricchio}, \bibinfo{person}{Marco Bertini}, \bibinfo{person}{Alberto
  Del~Bimbo}, {and} \bibinfo{person}{Shih-Fu Chang}.}
  \bibinfo{year}{2015}\natexlab{}.
\newblock \showarticletitle{Image popularity prediction in social media using
  sentiment and context features}. In \bibinfo{booktitle}{\emph{Proceedings of
  the ACM International Conference on Multimedia (ACMMM)}}.
  \bibinfo{pages}{907--910}.
\newblock


\bibitem[\protect\citeauthoryear{Gemmell, Schimoler, Ramezani, Christiansen,
  and Mobasher}{Gemmell et~al\mbox{.}}{2009}]%
        {Gemmell09}
\bibfield{author}{\bibinfo{person}{Jonathan Gemmell}, \bibinfo{person}{Thomas
  Schimoler}, \bibinfo{person}{Maryam Ramezani}, \bibinfo{person}{Laura
  Christiansen}, {and} \bibinfo{person}{Bamshad Mobasher}.}
  \bibinfo{year}{2009}\natexlab{}.
\newblock \showarticletitle{Improving Folkrank with item-based collaborative
  filtering}. In \bibinfo{booktitle}{\emph{Proceedings of the ACM RecSys
  Workshop on on Recommender Systems and the Social Web (RSWEB)}}.
  \bibinfo{pages}{17--24}.
\newblock


\bibitem[\protect\citeauthoryear{Godoy and Corbellini}{Godoy and
  Corbellini}{2016}]%
        {godoy2016folksonomy}
\bibfield{author}{\bibinfo{person}{Daniela Godoy} {and}
  \bibinfo{person}{Alejandro Corbellini}.} \bibinfo{year}{2016}\natexlab{}.
\newblock \showarticletitle{Folksonomy-based recommender systems: A
  state-of-the-art review}.
\newblock \bibinfo{journal}{\emph{International Journal of Intelligent
  Systems}} \bibinfo{volume}{31}, \bibinfo{number}{4} (\bibinfo{year}{2016}),
  \bibinfo{pages}{314--346}.
\newblock


\bibitem[\protect\citeauthoryear{Gu, Cai, Wang, and Chen}{Gu
  et~al\mbox{.}}{2018}]%
        {gu2017stack}
\bibfield{author}{\bibinfo{person}{Jiuxiang Gu}, \bibinfo{person}{Jianfei Cai},
  \bibinfo{person}{Gang Wang}, {and} \bibinfo{person}{Tsuhan Chen}.}
  \bibinfo{year}{2018}\natexlab{}.
\newblock \showarticletitle{Stack-captioning: Coarse-to-fine learning for image
  captioning}. In \bibinfo{booktitle}{\emph{Proceedings of the Thirty-Second
  AAAI Conference on Artificial Intelligence (AAAI)}}.
  \bibinfo{pages}{6837--6844}.
\newblock


\bibitem[\protect\citeauthoryear{Han, Wei, Cao, Yang, and Zhou}{Han
  et~al\mbox{.}}{2014}]%
        {Yahong14}
\bibfield{author}{\bibinfo{person}{Yahong Han}, \bibinfo{person}{Xingxing Wei},
  \bibinfo{person}{Xiaochun Cao}, \bibinfo{person}{Yi Yang}, {and}
  \bibinfo{person}{Xiaofang Zhou}.} \bibinfo{year}{2014}\natexlab{}.
\newblock \showarticletitle{Augmenting image descriptions using structured
  prediction output}.
\newblock \bibinfo{journal}{\emph{IEEE Transactions on Multimedia (TMM)}}
  \bibinfo{volume}{16}, \bibinfo{number}{6} (\bibinfo{year}{2014}),
  \bibinfo{pages}{1665--1676}.
\newblock


\bibitem[\protect\citeauthoryear{He, Gao, Kan, Liu, and Sugiyama}{He
  et~al\mbox{.}}{2014}]%
        {he2014predicting}
\bibfield{author}{\bibinfo{person}{Xiangnan He}, \bibinfo{person}{Ming Gao},
  \bibinfo{person}{Min-Yen Kan}, \bibinfo{person}{Yiqun Liu}, {and}
  \bibinfo{person}{Kazunari Sugiyama}.} \bibinfo{year}{2014}\natexlab{}.
\newblock \showarticletitle{Predicting the popularity of web 2.0 items based on
  user comments}. In \bibinfo{booktitle}{\emph{Proceedings of the 37th
  International ACM SIGIR Conference on Research \& Development in Information
  Retrieval}}. \bibinfo{pages}{233--242}.
\newblock


\bibitem[\protect\citeauthoryear{He, Liao, Zhang, Nie, Hu, and Chua}{He
  et~al\mbox{.}}{2017}]%
        {he2017neural}
\bibfield{author}{\bibinfo{person}{Xiangnan He}, \bibinfo{person}{Lizi Liao},
  \bibinfo{person}{Hanwang Zhang}, \bibinfo{person}{Liqiang Nie},
  \bibinfo{person}{Xia Hu}, {and} \bibinfo{person}{Tat-Seng Chua}.}
  \bibinfo{year}{2017}\natexlab{}.
\newblock \showarticletitle{Neural collaborative filtering}. In
  \bibinfo{booktitle}{\emph{Proceedings of the 26th International Conference on
  World Wide Web (WWW)}}. \bibinfo{pages}{173--182}.
\newblock


\bibitem[\protect\citeauthoryear{He, Zhang, Kan, and Chua}{He
  et~al\mbox{.}}{2016}]%
        {he2016fast}
\bibfield{author}{\bibinfo{person}{Xiangnan He}, \bibinfo{person}{Hanwang
  Zhang}, \bibinfo{person}{Min-Yen Kan}, {and} \bibinfo{person}{Tat-Seng
  Chua}.} \bibinfo{year}{2016}\natexlab{}.
\newblock \showarticletitle{Fast matrix factorization for online recommendation
  with implicit feedback}. In \bibinfo{booktitle}{\emph{Proceedings of the 39th
  International ACM SIGIR conference on Research and Development in Information
  Retrieval}}. \bibinfo{pages}{549--558}.
\newblock


\bibitem[\protect\citeauthoryear{Hotho, Jaschke, Schmitz, and Stumme}{Hotho
  et~al\mbox{.}}{2006}]%
        {hotho2006folkrank}
\bibfield{author}{\bibinfo{person}{Andreas Hotho}, \bibinfo{person}{Robert
  Jaschke}, \bibinfo{person}{Christoph Schmitz}, {and} \bibinfo{person}{Gerd
  Stumme}.} \bibinfo{year}{2006}\natexlab{}.
\newblock \showarticletitle{Folkrank: A ranking algorithm for folksonomies}. In
  \bibinfo{booktitle}{\emph{Proceedings of the LWA}}, Vol.~\bibinfo{volume}{1}.
  \bibinfo{pages}{111--114}.
\newblock


\bibitem[\protect\citeauthoryear{Hu, Yamasaki, and Aizawa}{Hu
  et~al\mbox{.}}{2016}]%
        {Hu2016}
\bibfield{author}{\bibinfo{person}{Jiani Hu}, \bibinfo{person}{Toshihiko
  Yamasaki}, {and} \bibinfo{person}{Kiyoharu Aizawa}.}
  \bibinfo{year}{2016}\natexlab{}.
\newblock \showarticletitle{Multimodal learning for image popularity prediction
  on social media}. In \bibinfo{booktitle}{\emph{Proceedings of the IEEE
  International Conference on Consumer Electronics - Taiwan (ICCE-TW)}}.
  \bibinfo{pages}{264--265}.
\newblock


\bibitem[\protect\citeauthoryear{Huang, Chen, Lin, Kang, and Yang}{Huang
  et~al\mbox{.}}{2018}]%
        {huang2018random}
\bibfield{author}{\bibinfo{person}{Feitao Huang}, \bibinfo{person}{Junhong
  Chen}, \bibinfo{person}{Zehang Lin}, \bibinfo{person}{Peipei Kang}, {and}
  \bibinfo{person}{Zhenguo Yang}.} \bibinfo{year}{2018}\natexlab{}.
\newblock \showarticletitle{Random forest exploiting post-related and
  user-related features for social media popularity prediction}. In
  \bibinfo{booktitle}{\emph{Proceedings of the ACM International Conference on
  Multimedia (ACMMM)}}. \bibinfo{pages}{2013--2017}.
\newblock


\bibitem[\protect\citeauthoryear{Jabeen, Khusro, Majid, and Rauf}{Jabeen
  et~al\mbox{.}}{2016}]%
        {Jabeen2016}
\bibfield{author}{\bibinfo{person}{Fouzia Jabeen}, \bibinfo{person}{Shah
  Khusro}, \bibinfo{person}{Amna Majid}, {and} \bibinfo{person}{Azhar Rauf}.}
  \bibinfo{year}{2016}\natexlab{}.
\newblock \showarticletitle{Semantics discovery in social tagging systems: A
  review}.
\newblock \bibinfo{journal}{\emph{Multimedia Tools and Applications}}
  \bibinfo{volume}{75}, \bibinfo{number}{1} (\bibinfo{year}{2016}),
  \bibinfo{pages}{573--605}.
\newblock


\bibitem[\protect\citeauthoryear{Jing, Su, Nie, Bai, Liu, and Wang}{Jing
  et~al\mbox{.}}{2018}]%
        {jing2018low}
\bibfield{author}{\bibinfo{person}{Peiguang Jing}, \bibinfo{person}{Yuting Su},
  \bibinfo{person}{Liqiang Nie}, \bibinfo{person}{Xu Bai},
  \bibinfo{person}{Jing Liu}, {and} \bibinfo{person}{Meng Wang}.}
  \bibinfo{year}{2018}\natexlab{}.
\newblock \showarticletitle{Low-rank multi-view embedding learning for
  micro-video popularity prediction}.
\newblock \bibinfo{journal}{\emph{IEEE Transactions on Knowledge and Data
  Engineering}} \bibinfo{volume}{30}, \bibinfo{number}{8}
  (\bibinfo{year}{2018}), \bibinfo{pages}{1519--1532}.
\newblock


\bibitem[\protect\citeauthoryear{Konstan, Miller, Maltz, Herlocker, Gordon, and
  Riedl}{Konstan et~al\mbox{.}}{1997}]%
        {konstan1997grouplens}
\bibfield{author}{\bibinfo{person}{Joseph~A Konstan},
  \bibinfo{person}{Bradley~N Miller}, \bibinfo{person}{David Maltz},
  \bibinfo{person}{Jonathan~L Herlocker}, \bibinfo{person}{Lee~R Gordon}, {and}
  \bibinfo{person}{John Riedl}.} \bibinfo{year}{1997}\natexlab{}.
\newblock \showarticletitle{GroupLens: applying collaborative filtering to
  Usenet news}.
\newblock \bibinfo{journal}{\emph{Commun. ACM}} \bibinfo{volume}{40},
  \bibinfo{number}{3} (\bibinfo{year}{1997}), \bibinfo{pages}{77--87}.
\newblock


\bibitem[\protect\citeauthoryear{Kulkarni, Premraj, Ordonez, Dhar, Li, Choi,
  Berg, and Berg}{Kulkarni et~al\mbox{.}}{2013}]%
        {kulkarni2013babytalk}
\bibfield{author}{\bibinfo{person}{Girish Kulkarni}, \bibinfo{person}{Visruth
  Premraj}, \bibinfo{person}{Vicente Ordonez}, \bibinfo{person}{Sagnik Dhar},
  \bibinfo{person}{Siming Li}, \bibinfo{person}{Yejin Choi},
  \bibinfo{person}{Alexander~C Berg}, {and} \bibinfo{person}{Tamara~L Berg}.}
  \bibinfo{year}{2013}\natexlab{}.
\newblock \showarticletitle{Babytalk: Understanding and generating simple image
  descriptions}.
\newblock \bibinfo{journal}{\emph{IEEE Transactions on Pattern Analysis and
  Machine Intelligence}} \bibinfo{volume}{35}, \bibinfo{number}{12}
  (\bibinfo{year}{2013}), \bibinfo{pages}{2891--2903}.
\newblock


\bibitem[\protect\citeauthoryear{Landia, Anand, Hotho, Jaschke, Doerfel, and
  Mitzlaff}{Landia et~al\mbox{.}}{2012}]%
        {landia2012extending}
\bibfield{author}{\bibinfo{person}{Nikolas Landia},
  \bibinfo{person}{Sarabjot~Singh Anand}, \bibinfo{person}{Andreas Hotho},
  \bibinfo{person}{Robert Jaschke}, \bibinfo{person}{Stephan Doerfel}, {and}
  \bibinfo{person}{Folke Mitzlaff}.} \bibinfo{year}{2012}\natexlab{}.
\newblock \showarticletitle{Extending FolkRank with content data}. In
  \bibinfo{booktitle}{\emph{Proceedings of the 4th ACM RecSys workshop on
  Recommender Systems and the Social Web}}. \bibinfo{pages}{1--8}.
\newblock


\bibitem[\protect\citeauthoryear{Le~Callet, M{\"o}ller, Perkis,
  et~al\mbox{.}}{Le~Callet et~al\mbox{.}}{2012}]%
        {le2012qualinet}
\bibfield{author}{\bibinfo{person}{Patrick Le~Callet},
  \bibinfo{person}{Sebastian M{\"o}ller}, \bibinfo{person}{Andrew Perkis},
  {et~al\mbox{.}}} \bibinfo{year}{2012}\natexlab{}.
\newblock \showarticletitle{Qualinet white paper on definitions of quality of
  experience}.
\newblock \bibinfo{journal}{\emph{European network on quality of experience in
  multimedia systems and services (COST Action IC 1003)}}  \bibinfo{volume}{3}
  (\bibinfo{year}{2012}).
\newblock


\bibitem[\protect\citeauthoryear{Lee, Chen, Hua, Hu, and He}{Lee
  et~al\mbox{.}}{2018}]%
        {10.1007/978-3-030-01225-0_13}
\bibfield{author}{\bibinfo{person}{Kuang-Huei Lee}, \bibinfo{person}{Xi Chen},
  \bibinfo{person}{Gang Hua}, \bibinfo{person}{Houdong Hu}, {and}
  \bibinfo{person}{Xiaodong He}.} \bibinfo{year}{2018}\natexlab{}.
\newblock \showarticletitle{Stacked Cross Attention for Image-Text Matching}.
  In \bibinfo{booktitle}{\emph{Proceedings of the European Conference on
  Computer Vision (ECCV)}}. \bibinfo{pages}{212--228}.
\newblock


\bibitem[\protect\citeauthoryear{Lin, Huang, Li, Yang, and Liu}{Lin
  et~al\mbox{.}}{2018}]%
        {lin2018layer}
\bibfield{author}{\bibinfo{person}{Zehang Lin}, \bibinfo{person}{Feitao Huang},
  \bibinfo{person}{Yukun Li}, \bibinfo{person}{Zhenguo Yang}, {and}
  \bibinfo{person}{Wenyin Liu}.} \bibinfo{year}{2018}\natexlab{}.
\newblock \showarticletitle{A layer-wise deep stacking model for social image
  popularity prediction}.
\newblock \bibinfo{journal}{\emph{Journal of World Wide Web}}
  \bibinfo{volume}{22}, \bibinfo{number}{4} (\bibinfo{year}{2018}),
  \bibinfo{pages}{1--17}.
\newblock


\bibitem[\protect\citeauthoryear{Lops, de~Gemmis, Semeraro, Musto, and
  Narducci}{Lops et~al\mbox{.}}{2013}]%
        {Lops13}
\bibfield{author}{\bibinfo{person}{Pasquale Lops}, \bibinfo{person}{Marco de
  Gemmis}, \bibinfo{person}{Giovanni Semeraro}, \bibinfo{person}{Cataldo
  Musto}, {and} \bibinfo{person}{Fedelucio Narducci}.}
  \bibinfo{year}{2013}\natexlab{}.
\newblock \showarticletitle{Content-based and collaborative techniques for tag
  recommendation: an empirical evaluation}.
\newblock \bibinfo{journal}{\emph{Journal of Intelligent Information Systems}}
  \bibinfo{volume}{40}, \bibinfo{number}{1} (\bibinfo{year}{2013}),
  \bibinfo{pages}{41--61}.
\newblock


\bibitem[\protect\citeauthoryear{Ma, Zhou, Tang, Tian, Al-Dhelaan, Al-Rodhaan,
  and Lee}{Ma et~al\mbox{.}}{2015}]%
        {ma2015social}
\bibfield{author}{\bibinfo{person}{Tinghuai Ma}, \bibinfo{person}{Jinjuan
  Zhou}, \bibinfo{person}{Meili Tang}, \bibinfo{person}{Yuan Tian},
  \bibinfo{person}{Abdullah Al-Dhelaan}, \bibinfo{person}{Mznah Al-Rodhaan},
  {and} \bibinfo{person}{Sungyoung Lee}.} \bibinfo{year}{2015}\natexlab{}.
\newblock \showarticletitle{Social network and tag sources based augmenting
  collaborative recommender system}.
\newblock \bibinfo{journal}{\emph{IEICE Transactions on Information and
  Systems}} \bibinfo{volume}{98}, \bibinfo{number}{4} (\bibinfo{year}{2015}),
  \bibinfo{pages}{902--910}.
\newblock


\bibitem[\protect\citeauthoryear{Massip, Hidayati, Cheng, and Hua}{Massip
  et~al\mbox{.}}{2018}]%
        {massip2018exploiting}
\bibfield{author}{\bibinfo{person}{Eric Massip},
  \bibinfo{person}{Shintami~Chusnul Hidayati}, \bibinfo{person}{Wen-Huang
  Cheng}, {and} \bibinfo{person}{Kai-Lung Hua}.}
  \bibinfo{year}{2018}\natexlab{}.
\newblock \showarticletitle{Exploiting category-specific information for image
  popularity prediction in social media}. In
  \bibinfo{booktitle}{\emph{Proceedings of the IEEE International Conference on
  Multimedia \& Expo Workshops (ICMEW)}}. \bibinfo{pages}{45--46}.
\newblock


\bibitem[\protect\citeauthoryear{McParlane, Moshfeghi, and Jose}{McParlane
  et~al\mbox{.}}{2014}]%
        {McParlane:2014:NCH:2578726.2578776}
\bibfield{author}{\bibinfo{person}{Philip~J. McParlane},
  \bibinfo{person}{Yashar Moshfeghi}, {and} \bibinfo{person}{Joemon~M. Jose}.}
  \bibinfo{year}{2014}\natexlab{}.
\newblock \showarticletitle{"Nobody comes here anymore, It's too crowded";
  Predicting image popularity on Flickr}. In
  \bibinfo{booktitle}{\emph{Proceedings of the ACM International Conference on
  Multimedia Retrieval (ICMR)}}. \bibinfo{pages}{385:385--385:391}.
\newblock


\bibitem[\protect\citeauthoryear{Meghawat, Yadav, Mahata, Yin, Shah, and
  Zimmermann}{Meghawat et~al\mbox{.}}{2018}]%
        {meghawat2018multimodal}
\bibfield{author}{\bibinfo{person}{Mayank Meghawat}, \bibinfo{person}{Satyendra
  Yadav}, \bibinfo{person}{Debanjan Mahata}, \bibinfo{person}{Yifang Yin},
  \bibinfo{person}{Rajiv~Ratn Shah}, {and} \bibinfo{person}{Roger Zimmermann}.}
  \bibinfo{year}{2018}\natexlab{}.
\newblock \showarticletitle{A multimodal approach to predict social media
  popularity}. In \bibinfo{booktitle}{\emph{Proceedings of the IEEE Conference
  on Multimedia Information Processing and Retrieval (MIPR)}}.
  \bibinfo{pages}{190--195}.
\newblock


\bibitem[\protect\citeauthoryear{Mishra, Rizoiu, and Xie}{Mishra
  et~al\mbox{.}}{2018}]%
        {DBLP:conf/icwsm/MishraRX18}
\bibfield{author}{\bibinfo{person}{Swapnil Mishra},
  \bibinfo{person}{Marian{-}Andrei Rizoiu}, {and} \bibinfo{person}{Lexing
  Xie}.} \bibinfo{year}{2018}\natexlab{}.
\newblock \showarticletitle{Modeling Popularity in Asynchronous Social Media
  Streams with Recurrent Neural Networks}. In
  \bibinfo{booktitle}{\emph{Proceedings of the Twelfth International Conference
  on Web and Social Media (ICWSM)}}. \bibinfo{pages}{201--210}.
\newblock


\bibitem[\protect\citeauthoryear{Nguyen, Wistuba, Grabocka, Drumond, and
  Schmidt-Thieme}{Nguyen et~al\mbox{.}}{2017}]%
        {10.1007/978-3-319-57454-7_15}
\bibfield{author}{\bibinfo{person}{Hanh T.~H. Nguyen}, \bibinfo{person}{Martin
  Wistuba}, \bibinfo{person}{Josif Grabocka}, \bibinfo{person}{Lucas~Rego
  Drumond}, {and} \bibinfo{person}{Lars Schmidt-Thieme}.}
  \bibinfo{year}{2017}\natexlab{}.
\newblock \showarticletitle{Personalized Deep Learning for Tag Recommendation}.
  In \bibinfo{booktitle}{\emph{Proceedings of the Pacific-Asia Conference on
  knowledge discovery and data mining}}. \bibinfo{pages}{186--197}.
\newblock


\bibitem[\protect\citeauthoryear{Nwana, Avestimehr, and Chen}{Nwana
  et~al\mbox{.}}{2013}]%
        {Nwana2013ALS}
\bibfield{author}{\bibinfo{person}{Amandianeze~O. Nwana},
  \bibinfo{person}{Amir~Salman Avestimehr}, {and} \bibinfo{person}{Tsuhan
  Chen}.} \bibinfo{year}{2013}\natexlab{}.
\newblock \showarticletitle{A latent social approach to YouTube popularity
  prediction}. In \bibinfo{booktitle}{\emph{Proceedings of the IEEE Global
  Communications Conference (GLOBECOM)}}. \bibinfo{pages}{3138--3144}.
\newblock


\bibitem[\protect\citeauthoryear{Page, Brin, Motwani, and Winograd}{Page
  et~al\mbox{.}}{1999}]%
        {ilprints422}
\bibfield{author}{\bibinfo{person}{Lawrence Page}, \bibinfo{person}{Sergey
  Brin}, \bibinfo{person}{Rajeev Motwani}, {and} \bibinfo{person}{Terry
  Winograd}.} \bibinfo{year}{1999}\natexlab{}.
\newblock \bibinfo{booktitle}{\emph{The PageRank Citation Ranking: Bringing
  Order to the Web.}}
\newblock \bibinfo{type}{Technical Report} 1999-66.
  \bibinfo{institution}{Stanford Digital Library Technologies Project}.
\newblock
\urldef\tempurl%
\url{http://ilpubs.stanford.edu:8090/422/}
\showURL{%
\tempurl}


\bibitem[\protect\citeauthoryear{Rathord, Jain, and Agrawal}{Rathord
  et~al\mbox{.}}{2019}]%
        {rathord2019comprehensive}
\bibfield{author}{\bibinfo{person}{Priyanka Rathord}, \bibinfo{person}{Anurag
  Jain}, {and} \bibinfo{person}{Chetan Agrawal}.}
  \bibinfo{year}{2019}\natexlab{}.
\newblock \showarticletitle{A comprehensive review on online news popularity
  prediction using machine learning approach}.
\newblock \bibinfo{journal}{\emph{International Journal Online of Science}}
  \bibinfo{volume}{5}, \bibinfo{number}{1} (\bibinfo{year}{2019}),
  \bibinfo{pages}{7--7}.
\newblock


\bibitem[\protect\citeauthoryear{Resnick, Iacovou, Suchak, Bergstrom, and
  Riedl}{Resnick et~al\mbox{.}}{1994}]%
        {resnick1994grouplens}
\bibfield{author}{\bibinfo{person}{Paul Resnick}, \bibinfo{person}{Neophytos
  Iacovou}, \bibinfo{person}{Mitesh Suchak}, \bibinfo{person}{Peter Bergstrom},
  {and} \bibinfo{person}{John Riedl}.} \bibinfo{year}{1994}\natexlab{}.
\newblock \showarticletitle{GroupLens: an open architecture for collaborative
  filtering of netnews}. In \bibinfo{booktitle}{\emph{Proceedings of the ACM
  conference on Computer Supported Cooperative Work}}.
  \bibinfo{pages}{175--186}.
\newblock


\bibitem[\protect\citeauthoryear{Rezaeenour, Eili, Hadavandi, and
  Roozbahani}{Rezaeenour et~al\mbox{.}}{2018}]%
        {rezaeenour2018developing}
\bibfield{author}{\bibinfo{person}{Jalal Rezaeenour},
  \bibinfo{person}{Mansoureh~Yari Eili}, \bibinfo{person}{Esmaeil Hadavandi},
  {and} \bibinfo{person}{Mohammad~Hossein Roozbahani}.}
  \bibinfo{year}{2018}\natexlab{}.
\newblock \showarticletitle{Developing a new hybrid intelligent approach for
  prediction online news popularity}.
\newblock \bibinfo{journal}{\emph{International Journal of Information Science
  and Management (IJISM)}} \bibinfo{volume}{16}, \bibinfo{number}{1}
  (\bibinfo{year}{2018}), \bibinfo{pages}{71--87}.
\newblock


\bibitem[\protect\citeauthoryear{Santosh, De~Sarkar, and Mukherjee}{Santosh
  et~al\mbox{.}}{2018}]%
        {santosh2018product}
\bibfield{author}{\bibinfo{person}{KC Santosh}, \bibinfo{person}{Sohan
  De~Sarkar}, {and} \bibinfo{person}{Arjun Mukherjee}.}
  \bibinfo{year}{2018}\natexlab{}.
\newblock \showarticletitle{Product Popularity Modeling Via Time Series
  Embedding}. In \bibinfo{booktitle}{\emph{Proceedings of the IEEE/ACM
  International Conference on Advances in Social Networks Analysis and Mining
  (ASONAM)}}. \bibinfo{pages}{650--653}.
\newblock


\bibitem[\protect\citeauthoryear{Si, Liu, Li, Jiang, and Sun}{Si
  et~al\mbox{.}}{2009}]%
        {Si09}
\bibfield{author}{\bibinfo{person}{Xiance Si}, \bibinfo{person}{Zhiyuan Liu},
  \bibinfo{person}{Peng Li}, \bibinfo{person}{Qixia Jiang}, {and}
  \bibinfo{person}{Maosong Sun}.} \bibinfo{year}{2009}\natexlab{}.
\newblock \showarticletitle{Content-based and graph-based tag suggestion}. In
  \bibinfo{booktitle}{\emph{Proceedings of the European Conference on Machine
  Learning and Principles and Practice of Knowledge Discovery in Databases
  (ECMLPKDD) Discovery Challenge}}. \bibinfo{pages}{243--260}.
\newblock


\bibitem[\protect\citeauthoryear{Sigurbj\"{o}rnsson and van
  Zwol}{Sigurbj\"{o}rnsson and van Zwol}{2008}]%
        {Sigurbjornsson08}
\bibfield{author}{\bibinfo{person}{B. Sigurbj\"{o}rnsson} {and}
  \bibinfo{person}{R. van Zwol}.} \bibinfo{year}{2008}\natexlab{}.
\newblock \showarticletitle{Flickr tag recommendation based on collective
  knowledge}. In \bibinfo{booktitle}{\emph{Proceedings of the World Wide Web
  Conference on World Wide Web (WWW)}}. \bibinfo{pages}{327--336}.
\newblock


\bibitem[\protect\citeauthoryear{Singhal, Sinha, and Pant}{Singhal
  et~al\mbox{.}}{2017}]%
        {10.5120/ijca2017916055}
\bibfield{author}{\bibinfo{person}{Ayush Singhal}, \bibinfo{person}{Pradeep
  Sinha}, {and} \bibinfo{person}{Rakesh Pant}.}
  \bibinfo{year}{2017}\natexlab{}.
\newblock \showarticletitle{Use of Deep Learning in Modern Recommendation
  System: A Summary of Recent Works}.
\newblock \bibinfo{journal}{\emph{International Journal of Computer
  Applications}} \bibinfo{volume}{180}, \bibinfo{number}{7}
  (\bibinfo{year}{2017}), \bibinfo{pages}{17--22}.
\newblock


\bibitem[\protect\citeauthoryear{Su and Khoshgoftaar}{Su and
  Khoshgoftaar}{2009}]%
        {Su:2009:SCF:1592474.1722966}
\bibfield{author}{\bibinfo{person}{Xiaoyuan Su} {and} \bibinfo{person}{Taghi~M.
  Khoshgoftaar}.} \bibinfo{year}{2009}\natexlab{}.
\newblock \showarticletitle{A Survey of Collaborative Filtering Techniques}.
\newblock \bibinfo{journal}{\emph{Advances in Artificial Intelligence}}
  \bibinfo{volume}{2009} (\bibinfo{year}{2009}), \bibinfo{pages}{4:2--4:2}.
\newblock


\bibitem[\protect\citeauthoryear{Thomee, Shamma, Friedland, Elizalde, Ni,
  Poland, Borth, and Li}{Thomee et~al\mbox{.}}{2016}]%
        {thomee2016yfcc100m}
\bibfield{author}{\bibinfo{person}{Bart Thomee}, \bibinfo{person}{David~A.
  Shamma}, \bibinfo{person}{Gerald Friedland}, \bibinfo{person}{Benjamin
  Elizalde}, \bibinfo{person}{Karl Ni}, \bibinfo{person}{Douglas Poland},
  \bibinfo{person}{Damian Borth}, {and} \bibinfo{person}{Li-Jia Li}.}
  \bibinfo{year}{2016}\natexlab{}.
\newblock \showarticletitle{{YFCC100M}: The New Data in Multimedia Research}.
\newblock \bibinfo{journal}{\emph{Commun. ACM}} \bibinfo{volume}{59},
  \bibinfo{number}{2} (\bibinfo{year}{2016}), \bibinfo{pages}{64--73}.
\newblock


\bibitem[\protect\citeauthoryear{van Zwol, Rae, and Pueyo}{van Zwol
  et~al\mbox{.}}{2010}]%
        {vanZwol2010}
\bibfield{author}{\bibinfo{person}{Roelof van Zwol}, \bibinfo{person}{Adam
  Rae}, {and} \bibinfo{person}{Lluis~Garcia Pueyo}.}
  \bibinfo{year}{2010}\natexlab{}.
\newblock \showarticletitle{Prediction of favourite photos using social,
  visual, and textual signals}. In \bibinfo{booktitle}{\emph{Proceedings of the
  ACM International Conference on Multimedia (ACMMM)}}.
  \bibinfo{pages}{1015--1018}.
\newblock


\bibitem[\protect\citeauthoryear{Wang, Wang, and Yeung}{Wang
  et~al\mbox{.}}{2015}]%
        {Wang:2015:CDL:2783258.2783273}
\bibfield{author}{\bibinfo{person}{Hao Wang}, \bibinfo{person}{Naiyan Wang},
  {and} \bibinfo{person}{Dit-Yan Yeung}.} \bibinfo{year}{2015}\natexlab{}.
\newblock \showarticletitle{Collaborative deep learning for recommender
  systems}. In \bibinfo{booktitle}{\emph{Proceedings of the 21th ACM SIGKDD
  International Conference on Knowledge Discovery and Data Mining}}
  \emph{(\bibinfo{series}{KDD '15})}. \bibinfo{pages}{1235--1244}.
\newblock


\bibitem[\protect\citeauthoryear{Wang, Enokibori, Hirayama, Hara, and
  Mase}{Wang et~al\mbox{.}}{2017}]%
        {Wang:2017:UGB:3132515.3132523}
\bibfield{author}{\bibinfo{person}{Xueting Wang}, \bibinfo{person}{Yu
  Enokibori}, \bibinfo{person}{Takatsugu Hirayama}, \bibinfo{person}{Kensho
  Hara}, {and} \bibinfo{person}{Kenji Mase}.} \bibinfo{year}{2017}\natexlab{}.
\newblock \showarticletitle{User Group Based Viewpoint Recommendation Using
  User Attributes for Multiview Videos}. In
  \bibinfo{booktitle}{\emph{Proceedings of the Workshop on Multimodal
  Understanding of Social, Affective and Subjective Attributes}}
  \emph{(\bibinfo{series}{MUSA2 '17})}. \bibinfo{pages}{3--9}.
\newblock


\bibitem[\protect\citeauthoryear{Wu, Chen, Sun, Liu, Ghanem, and Lyu}{Wu
  et~al\mbox{.}}{2018}]%
        {Wu2018TaggingLH}
\bibfield{author}{\bibinfo{person}{Baoyuan Wu}, \bibinfo{person}{Weidong Chen},
  \bibinfo{person}{Peng Sun}, \bibinfo{person}{Weiwei Liu},
  \bibinfo{person}{Bernard Ghanem}, {and} \bibinfo{person}{Siwei Lyu}.}
  \bibinfo{year}{2018}\natexlab{}.
\newblock \showarticletitle{Tagging Like Humans: Diverse and Distinct Image
  Annotation}. In \bibinfo{booktitle}{\emph{Proceedings of the IEEE Conference
  on Computer Vision and Pattern Recognition (CVPR)}}.
  \bibinfo{pages}{7967--7975}.
\newblock


\bibitem[\protect\citeauthoryear{Wu, Cheng, Zhang, and Mei}{Wu
  et~al\mbox{.}}{2016}]%
        {Wu:2016}
\bibfield{author}{\bibinfo{person}{Bo Wu}, \bibinfo{person}{Wen-Huang Cheng},
  \bibinfo{person}{Yongdong Zhang}, {and} \bibinfo{person}{Tao Mei}.}
  \bibinfo{year}{2016}\natexlab{}.
\newblock \showarticletitle{Time Matters: Multi-scale temporalization of social
  media popularity}. In \bibinfo{booktitle}{\emph{Proceedings of the ACM
  International Conference on Multimedia (ACMMM)}}.
  \bibinfo{pages}{1336--1344}.
\newblock


\bibitem[\protect\citeauthoryear{Yamaguchi, Berg, and Ortiz}{Yamaguchi
  et~al\mbox{.}}{2014}]%
        {yamaguchi2014chic}
\bibfield{author}{\bibinfo{person}{Kota Yamaguchi}, \bibinfo{person}{Tamara~L
  Berg}, {and} \bibinfo{person}{Luis~E Ortiz}.}
  \bibinfo{year}{2014}\natexlab{}.
\newblock \showarticletitle{Chic or social: Visual popularity analysis in
  online fashion networks}. In \bibinfo{booktitle}{\emph{Proceedings of the
  22nd ACM International Conference on Multimedia (ACMMM)}}.
  \bibinfo{pages}{773--776}.
\newblock


\bibitem[\protect\citeauthoryear{Yamasaki, Hu, Sano, and Aizawa}{Yamasaki
  et~al\mbox{.}}{2017}]%
        {yamasaki2017folkpopularityrank}
\bibfield{author}{\bibinfo{person}{Toshihiko Yamasaki}, \bibinfo{person}{Jiani
  Hu}, \bibinfo{person}{Shumpei Sano}, {and} \bibinfo{person}{Kiyoharu
  Aizawa}.} \bibinfo{year}{2017}\natexlab{}.
\newblock \showarticletitle{Folkpopularityrank: Tag recommendation for
  enhancing social popularity using text tags in content sharing services}. In
  \bibinfo{booktitle}{\emph{Proceedings of the 26th International Joint
  Conference on Artificial Intelligence (IJCAI)}}. \bibinfo{pages}{3231--3237}.
\newblock


\bibitem[\protect\citeauthoryear{Yamasaki, Sano, and Aizawa}{Yamasaki
  et~al\mbox{.}}{2014}]%
        {Yamasaki14}
\bibfield{author}{\bibinfo{person}{Toshihiko Yamasaki},
  \bibinfo{person}{Shumpei Sano}, {and} \bibinfo{person}{Kiyoharu Aizawa}.}
  \bibinfo{year}{2014}\natexlab{}.
\newblock \showarticletitle{Social popularity score: Predicting numbers of
  views, comments, and favorites of social photos using only annotations}. In
  \bibinfo{booktitle}{\emph{Proceedings of the International Workshop on
  Internet-Scale Multimedia Management (WISMM)}}. \bibinfo{pages}{3--8}.
\newblock


\bibitem[\protect\citeauthoryear{Yu, Zhou, Deng, and Hu}{Yu
  et~al\mbox{.}}{2018}]%
        {yu2018tag}
\bibfield{author}{\bibinfo{person}{Hong Yu}, \bibinfo{person}{Bing Zhou},
  \bibinfo{person}{Mingyao Deng}, {and} \bibinfo{person}{Feng Hu}.}
  \bibinfo{year}{2018}\natexlab{}.
\newblock \showarticletitle{Tag recommendation method in folksonomy based on
  user tagging status}.
\newblock \bibinfo{journal}{\emph{Journal of Intelligent Information Systems}}
  (\bibinfo{year}{2018}), \bibinfo{pages}{1--22}.
\newblock


\bibitem[\protect\citeauthoryear{Zhang, Wang, Wang, and Zha}{Zhang
  et~al\mbox{.}}{2018}]%
        {zhang2018user}
\bibfield{author}{\bibinfo{person}{Wei Zhang}, \bibinfo{person}{Wen Wang},
  \bibinfo{person}{Jun Wang}, {and} \bibinfo{person}{Hongyuan Zha}.}
  \bibinfo{year}{2018}\natexlab{}.
\newblock \showarticletitle{User-guided hierarchical attention network for
  multi-modal social image popularity prediction}. In
  \bibinfo{booktitle}{\emph{Proceedings of the International Conference on
  World Wide Web (WWW)}}. \bibinfo{pages}{1277--1286}.
\newblock


\bibitem[\protect\citeauthoryear{Zhang, Gong, and Shah}{Zhang
  et~al\mbox{.}}{2016}]%
        {zhang2016fast}
\bibfield{author}{\bibinfo{person}{Yang Zhang}, \bibinfo{person}{Boqing Gong},
  {and} \bibinfo{person}{Mubarak Shah}.} \bibinfo{year}{2016}\natexlab{}.
\newblock \showarticletitle{Fast zero-shot image tagging}. In
  \bibinfo{booktitle}{\emph{Proceedings of the IEEE Conference on Computer
  Vision and Pattern Recognition (CVPR)}}. \bibinfo{pages}{5985--5994}.
\newblock


\bibitem[\protect\citeauthoryear{Zhang, Zhang, and Tang}{Zhang
  et~al\mbox{.}}{2009}]%
        {Zhang09}
\bibfield{author}{\bibinfo{person}{Yuan Zhang}, \bibinfo{person}{Ning Zhang},
  {and} \bibinfo{person}{Jie Tang}.} \bibinfo{year}{2009}\natexlab{}.
\newblock \showarticletitle{A collaborative filtering tag recommendation system
  based on graph}. In \bibinfo{booktitle}{\emph{Proceedings of the European
  Conference on Machine Learning and Principles and Practice of Knowledge
  Discovery in Databases (ECMLPKDD) Discovery Challenge}}.
  \bibinfo{pages}{297--306}.
\newblock


\bibitem[\protect\citeauthoryear{Zhu, Hanjalic, and Redi}{Zhu
  et~al\mbox{.}}{2016}]%
        {zhu2016qoe}
\bibfield{author}{\bibinfo{person}{Yi Zhu}, \bibinfo{person}{Alan Hanjalic},
  {and} \bibinfo{person}{Judith~A Redi}.} \bibinfo{year}{2016}\natexlab{}.
\newblock \showarticletitle{QoE prediction for enriched assessment of
  individual video viewing experience}. In
  \bibinfo{booktitle}{\emph{Proceedings of the 24th ACM International
  Conference on Multimedia (ACMMM)}}. \bibinfo{pages}{801--810}.
\newblock


\end{thebibliography}

\end{document}